\documentclass[aps,prd,twocolumn,showpacs,preprintnumbers,nofootinbib,superscriptaddress,10pt]{revtex4-2}

\usepackage{amsmath,amssymb}
\usepackage{graphicx}
\usepackage[colorlinks=true,citecolor=blue,linkcolor=blue,urlcolor=blue]{hyperref}
\usepackage{physics}
\usepackage{slashed}
\usepackage{color}
\usepackage[dvipsnames]{xcolor}

\usepackage{hyperref}
\hypersetup{%
    colorlinks = true,
    linkcolor  = MidnightBlue,
    urlcolor   = BrickRed,
    citecolor  = MidnightBlue
}%

\usepackage{caption}
\usepackage{subfig}

\def\nn{\nonumber}

\begin{document}

	\title{Constraining interacting dark energy models with black hole superradiance}

    \author{Zhen-Hong Lyu}
    \email{lyuzhenhong@itp.ac.cn}
    \affiliation{Institute of Theoretical Physics, Chinese Academy of Sciences (CAS), Beijing 100190, China}
    \affiliation{School of Physical Sciences, University of Chinese Academy of Sciences (UCAS), Beijing 100049, China}

    \author{Rong-Gen Cai}
    \email{caironggen@nbu.edu.cn}
    \affiliation{Institute of Fundamental Physics and Quantum Technology, \& School of Physical Science and Technology, Ningbo University, Ningbo, 315211, China}

    \author{Shao-Jiang Wang}
    \email{schwang@itp.ac.cn}
    \affiliation{Institute of Theoretical Physics, Chinese Academy of Sciences (CAS), Beijing 100190, China}
    \affiliation{Asia Pacific Center for Theoretical Physics (APCTP), Pohang 37673, Korea}

    \author{Xiang-Xi Zeng}
    \email{zengxiangxi@itp.ac.cn}
    \affiliation{Institute of Theoretical Physics, Chinese Academy of Sciences (CAS), Beijing 100190, China}
    \affiliation{School of Physical Sciences, University of Chinese Academy of Sciences (UCAS), Beijing 100049, China}

    \begin{abstract}
    The recent preference for a dynamical dark energy (DE) from the Dark Energy Spectroscopic Instrument seems to call for interactions between DE and dark matter (DM), either from direct DE-DM interaction or indirect interaction induced by modified gravity. Therefore, an independent probe for these kinds of DE-DM interactions would be appealing from observational aspects. In this paper, we propose the black hole superradiance as a novel astrophysical probe for field-theoretic interacting DE-DM models, providing complementary constraints independent of large-scale cosmological observations. The core principle is that the DE-DM interaction can alter the effective mass of the superradiant ultralight boson, thereby modifying its superradiant instability rate around spinning black holes. We explore this connection through two distinct scenarios: a model where the DE field mediates a dark fifth force within the DM sector, affecting the superradiance from DM particles; and a novel mechanism where the DE field itself becomes superradiant due to the effective mass enhancement induced by dense DM spikes around supermassive black holes. By applying a statistical framework to black hole observations in both scenarios, we derive constraints on the fundamental DE-DM coupling strength. Although the current constraints are rather loose due to small samples and inaccurate measurements, our work provides new astrophysical constraints on these interacting DE-DM scenarios and establishes a new synergy between black hole physics and cosmology for probing the fundamental nature of the dark sector.
    \end{abstract}

    \maketitle

\section{Introduction}~\label{sec:intro}

The $\Lambda$-cold-dark matter ($\Lambda$CDM) model, the standard cosmological framework, describes a Universe dominated by two dark components. The first is the dark matter (DM), a nonbaryonic substance introduced to explain phenomena such as galaxy rotation curves and a wide array of other astronomical observations~\cite{Zwicky1937,Rubin1978,Clowe2006,Bartelmann2001,Springel2005}. The second is the dark energy (DE), an exotic form of energy invoked to account for the observed accelerating expansion of the Universe~\cite{Frieman:2008sn}. Traditionally, DE is identified with the cosmological constant $\Lambda$, representing the vacuum energy that exerts a uniform negative pressure throughout space. However, this simple picture is troubled by the profound disagreement between its theoretically predicted and observationally inferred energy densities—a persistent challenge known as the cosmological constant problem~\cite{Weinberg1989}.

Despite this foundational issue, the $\Lambda$CDM model has been remarkably successful, largely supported by various observations in the era of precision cosmology~\cite{Planck2018}. Nevertheless, recent and increasingly precise measurements have revealed significant discrepancies. The most prominent one among these is the ``Hubble tension''~\cite{Cai:2023sli}, a persistent disagreement in the measured value of the Hubble constant, $H_0$, between early-Universe probes (like the Cosmic Microwave Background) and late-time, local Universe observations~\cite{Riess2022}. The other is the ``$S_8$ tension,'' relating to the differing values of the amplitude of matter fluctuations on cosmological scales as inferred from early- and late-Universe data~\cite{DES2022}. Another less discussed but non-negligible tension is the dubbed $\gamma$ tension~\cite{Nguyen:2023fip}, which might call for modified gravity effects. Furthermore, recent data from the Dark Energy Spectroscopic Instrument (DESI) have cast new doubt on the cosmological constant, showing a preference for dynamical DE models~\cite{DESI:2024mwx,DESI:2024aqx,DESI:2024kob}.
In light of these challenges, interacting dark energy (IDE) models have emerged as a compelling theoretical alternative. In these scenarios, DE and DM are not independently conserved but exchange energy through a nongravitational interaction~\cite{Wang:2016lxa,Pourtsidou:2013nha,Cai:2004dk}. Such models have long been proposed to address key cosmological puzzles, including the coincidence problem, the aforementioned $H_0$ and $S_8$ tensions, and the recent hints of dynamical DE from the DESI collaboration~\cite{Pourtsidou:2016ico,DiValentino:2017iww,Lucca:2020zjb,Cai:2021wgv,PhysRevLett.133.251003,Sabogal:2024yha,Zhai:2023yny,Bernui:2023byc,Wang:2025znm,Silva:2025hxw,Li:2024qso,Li:2025owk,Lee:2006za,Lee_2010}.

Typically, investigations into IDE models focus on their cosmological implications. By confronting theoretical predictions with data from the Cosmic Microwave Background and large-scale structure, the primary goal has been to understand the role of dark sector interactions in the overall cosmic expansion history. Our work, however, emphasizes a complementary perspective: the interaction within the dark sector can also profoundly impact physics on much smaller, astrophysical scales. We suggest that black hole (BH) superradiance provides a unique observational window to probe such small-scale effects. BH superradiance is a phenomenon where ultralight bosonic particles, interacting with a spinning BH, can trigger a rotational instability~\cite{zeldo1,Damour:1976kh,Teukolsky:1974yv,Press:1972zz} (see Ref.~\cite{Brito:2015oca} for a comprehensive review). This instability leads to the exponential growth of the boson field into a macroscopic ``cloud'' by rapidly extracting angular momentum from the BH. The potential observational consequences of BH superradiance have been widely explored, particularly the gravitational wave emission originating from the bosonic clouds~\cite{Arvanitaki:2014wva,Brito:2017wnc,Ghosh:2018gaw,LIGOScientific:2021rnv,Yuan:2022bem,Guo:2022mpr,liuGravitationalLaserStimulated2024,Kang:2024trj}, the signatures imprinted on binary systems~\cite{Baumann:2018vus,baumannGravitationalColliderPhysics2020,Baumann:2021fkf,Baumann:2022pkl,Berti:2019wnn,Zhang:2018kib,Zhang:2019eid,Ikeda:2020xvt,Boskovic:2024fga,Takahashi:2021yhy,Takahashi:2024fyq,Cao:2023fyv,Tong:2022bbl,Tomaselli:2023ysb,Tomaselli:2024dbw,Guo:2024iye,Guo:2025ckp,Peng:2025zca,Guo:2025pea,Li:2025qyu,Kyriazis:2025fis}, and the electromagnetic signals~\cite{rosaStimulatedAxionDecay2018,Ikeda:2018nhb,Spieksma:2023vwl,Chen:2021lvo,Chen:2022kzv,Bai:2025yxm}.

More pertinent to this work, BH superradiance has emerged as a powerful tool for probing fundamental physics beyond the Standard Model, especially within the dark sector~\cite{Arvanitaki:2009fg,Arvanitaki:2010sy,britoGravitationalWaveSearches2017,Stott:2018opm,Fernandez:2019qbj,Davoudiasl:2019nlo,Ng:2020ruv,Saha:2022hcd,chengConstrainsAxionlikeParticle2023,Witte:2024drg,Hoof:2024quk,Caputo:2025oap,Aswathi:2025nxa}. The core principle is that the observation of highly spinning BHs places stringent constraints on the existence of ultralight bosons within a characteristic mass range $\sim (G_NM)^{-1}$, where $M$ is the BH mass, and $G_N$ is the Newton’s constant (in natural units with $\hbar=c=1$). This makes superradiance an effective probe for new light particles near BHs. Such probing becomes particularly interesting in scenarios where the dark sector involves multiple interacting species~\cite{Fukuda:2019ewf,East:2022ppo,Zhu:2025enp,Lyu:2025lue} or is influenced by the cosmological background, such as the cosmic neutrino background~\cite{Lambiase:2025twn}.

Building upon this principle, our work extends the application of superradiance constraints to IDE models. While the aforementioned studies often focus on constraining the properties of a single ultralight boson (which may or may not constitute dark matter), we investigate how the coupling between DE and DM modifies the superradiance phenomenon, thereby allowing us to constrain the interaction itself. This unique sensitivity offers a robust way to test specific field-theoretic IDE models, particularly those where DE or DM consists of ultralight scalar fields susceptible to superradiance.

The work is structured as follows: In Section~\ref{sec:IDE}, we briefly review the theoretical foundations of IDE models, distinguishing between phenomenological and field-theoretic approaches. In Section~\ref{sec:sr}, we establish our primary constraint methodology, detailing the physics of BH superradiance and the statistical framework used to constrain model parameters. We then apply this methodology to our specific scenarios. In Section~\ref{sec:DEmediator}, we present the Model I, where the DE scalar field acts as a mediator for a dark fifth force within the DM sector, and derive the corresponding constraints. In Section~\ref{sec:DESR}, we construct and analyze the Model II, a more intricate scenario where the DE field itself exhibits a superradiant instability induced by a DM spike. Finally, we conclude this work and discuss possible future directions in Section~\ref{sec:conclusion}. Throughout this work, we adopt natural units with $\hbar=c=1$.

\section{Interacting dark energy model}~\label{sec:IDE}

Investigations into IDE models predominantly follow two distinct methodologies: the phenomenological parameterization approach and the more fundamental field-theoretic model approach.
    
\subsection{Parameterization approach}

In many cosmological studies, the dark components are treated as two separate, perfect fluids that permeate the Universe. In the context of an IDE, the standard assumption of individual energy conservation for DM and DE is relaxed. Instead, a nongravitational energy exchange between them is permitted. This interaction is phenomenologically characterized by an energy transfer rate, $Q$, which modifies the conservation equations for both dark sector components as~\cite{Wang:2016lxa}
\begin{align}
    &\dot{\rho}_{\text{dm}} + 3H \rho_{\text{dm}} = Q,\\
    &\dot{\rho}_{\text{de}} + 3H (1 + w) \rho_{\text{de}} = -Q.\label{fluid}
\end{align}
Here, $\rho_{\text{dm}}$ and $\rho_{\text{de}}$ are the energy densities of DM and DE, respectively, $H$ is the Hubble parameter, and $w$ is the equation of state parameter for DE. The sign of the interaction term $Q$ dictates the direction of energy flow: $Q > 0$ signifies energy transfer from DE to DM, while $Q < 0$ indicates the opposite direction.

Within the phenomenological framework, the interaction term $Q$ is parameterized based on general physical considerations rather than derived from a specific fundamental theory. Its functional form is generally constructed to be proportional to the energy densities of the dark components and the Hubble rate, $H$, for dimensional consistency. Common linear interaction kernels in the literature include $Q = 3H\xi\rho_{\text{dm}}$, $Q = 3H\xi\rho_{\text{de}}$, or a combination of both~\cite{DiValentino:2017iww,Lucca:2020zjb,Zhai:2023yny,Bernui:2023byc,vanderWesthuizen:2025vcb}. In these expressions, $\xi$ represents a dimensionless coupling constant that quantifies the strength of the interaction. This parameterized approach provides a convenient framework to study the cosmological consequences of various energy exchange scenarios and to constrain the strength of such interactions with observational data.
	
\subsection{Field-theoretic model}~\label{subsec:fieldtheoretic}

An alternative to the phenomenological fluid approach is to build models from a more fundamental, microscopic perspective, for example, an action with a generic Lagrangian. For illustration, we consider a coupled two-field model with the action given by
\begin{align}
S & = \int \mathrm{d}^4 x \, \sqrt{-g}\left[\frac{R}{16\pi G_N} + \mathcal{L}_{\phi}\bigl(\phi, \partial_{\mu} \phi\bigr)\right. \nn \\
  & \left.+ \mathcal{L}_{\chi}\bigl(\chi, \partial_{\mu} \chi\bigr) + \mathcal{L}_{\mathrm{int}}(\phi, \chi)\right] + S_{\mathrm{M}}\,.
\end{align}
The system contains two ultralight fields: DM $\chi$ and DE $\phi$. $S_\mathrm{M}$ represents the action for
the remaining Standard Model particles, such as radiation and baryonic fluids. The individual Lagrangians of $\phi$ and $\chi$ are defined by
\begin{align}
\mathcal{L}_{\phi}\left(\phi, \nabla_{\mu} \phi\right) & =-\frac{1}{2}\nabla_{\mu} \phi \nabla^{\mu} \phi-V(\phi),\\
\mathcal{L}_{\chi}\left(\chi, \nabla_{\mu} \chi\right) & =-\frac{1}{2}\nabla_{\mu} \chi \nabla^{\mu} \chi-U(\chi),\\
\mathcal{L}_{\mathrm{int}}(\phi, \chi) & =-W(\phi, \chi)\,.
\end{align}
Here $V(\phi)$ and $U(\chi)$ are the self-interaction potentials. The interaction between the dark components is explicitly defined by an interaction term $W(\phi,\chi)$. From this Lagrangian, the equations of motion for the fields $\phi$ and $\chi$ can be derived. By defining the energy density and pressure for each component appropriately (see Appendix \ref{app:Q_derivation} for details), these field equations can be recast into the familiar form of fluid continuity equations,
\begin{align}
\dot{\rho}_{\phi}+3 H\left(\rho_{\phi}+P_{\phi}\right) & =Q_\phi, \\
\dot{\rho}_{\chi}+3 H\left(\rho_{\chi}+P_{\chi}\right) & =Q_\chi, \\
\dot{\rho}_{\mathrm{M}}+3 H\left(\rho_{\mathrm{M}}+P_{\mathrm{M}}\right) & =0\,.
\end{align}

This field-theoretic procedure provides a direct link between the macroscopic cosmological evolution and the microscopic properties of the dark sectors, thereby establishing a physical origin for the phenomenological term $Q$ in Eq.~\eqref{fluid} (see also~\cite{Aboubrahim:2024cyk,Aboubrahim:2024spa,Johnson:2020gzn,Johnson:2021wou,Bansal:2024bbb}). As detailed in Appendix~\ref{app:Q_derivation}, if the individual energy densities are defined such that their sum equals the total energy density of the dark sector, the fluid continuity equations are recovered with interaction terms that satisfy $Q_\phi = -Q_\chi$. For the specific case where the interaction potential $W(\phi, \chi)$ is assigned entirely to the dark matter component, the energy transfer rate is explicitly found to be $Q_\chi = -Q_\phi = W_{,\phi}\dot{\phi}$.

The general two-scalar model presented above serves as an illustrative framework to demonstrate how a microscopic field theory can give rise to an effective interaction consistent with the phenomenological fluid description. Our work builds upon this field-theoretic approach by constructing and analyzing two specific models derived from fundamental Lagrangians, which will be detailed in Sections~\ref{sec:DEmediator} and \ref{sec:DESR}, respectively. Before turning to these models, we delve into the constraints from BH superradiance in Section~\ref{sec:sr}.

Our first model (Model I, Section~\ref{sec:DEmediator}) is a direct realization of the framework discussed here, employing a trilinear coupling ($W(\phi, \chi) \propto \phi\chi^2$) between minimally coupled scalar fields representing DE and DM. This represents a scenario of direct interaction mediated by a potential term.

Our second model (Model II, Section~\ref{sec:DESR}) explores a different mechanism where the interaction arises indirectly through a nonminimal coupling to gravity. In this scenario, the DE field $\Phi$ couples conformally to the DM sector via the action term $S_{\mathrm{DM}}(\Psi ; \Omega^{2}(\Phi) g_{\mu \nu})$. This coupling modifies the dynamics of the DE field itself, making its effective properties dependent on the local DM environment, without a direct potential term. This nonminimal coupling provides an alternative physical basis for interaction within the dark sector. Both models illustrate how macroscopic cosmic evolution links to microscopic dark sector particle properties, laying the groundwork for the later superradiance analysis.

\section{Constraints from black hole superradiance}~\label{sec:sr}

Having introduced the concept of IDE models, we now turn to an astrophysical method for constraining them. The BH superradiance, a phenomenon sensitive to the existence of ultralight bosonic fields, provides a unique observational window. Many field-theoretic IDE models naturally incorporate such ultralight scalars, playing the role of either DE or DM. Since their properties can be modified by the dark sector interaction, the BH superradiance can thus serve as a characteristic probe. In this section, we will first provide a brief overview of BH superradiance and then detail the methodology used to translate superradiance constraints into exclusion limits on the parameter space of relevant IDE models.

\subsection{Superradiance and black hole evolution}

The interaction between a rotating BH and an ultralight scalar field $\phi$ with mass $\mu$ can trigger instability of the wave equation $(\square-\mu^2) \phi=0$.  This superradiant instability arises due to the presence of growing modes for $\phi$, which can be identified as the imaginary part of the eigenvalues in the associated Schr\"{o}dinger equation. The superradiance rate depends on $\mu$ and is approximately maximized when the dimensionless ``gravitational coupling'' of the atomlike bosonic states around a spinning BH~\cite{Baumann:2019eav},
\begin{equation}
\alpha\equiv G_N M \mu\approx 0.08\left(\frac{M}{10 \mathrm{M}_{\odot}}\right)\left(\frac{\mu}{10^{-12} \mathrm{eV}}\right),
\end{equation}
is $\alpha \sim \mathcal{O}(0.1)$.

The bosonic state around the BH can be labelled by ``quantum numbers'' $(n\ell m)$ and has the complex-valued frequency $\omega=\omega_R+i\omega_I$. The $\ket{n\ell m}$ level is superradiant, i.e., $\omega_I>0$, if the superradiance condition 
\begin{equation}
\frac{\alpha}{m}<\frac{\tilde{a}}{2\left(1+\sqrt{1-\tilde{a}^{2}}\right)}
\end{equation}
is fulfilled. Here, $\tilde{a}=J/M^2$ is the dimensionless spin parameter of BH. In the nonrelativistic approximation, the real component of the angular frequency is~\cite{Baumann:2019eav}
\begin{align}
\omega^{R}_{n \ell m}
    = \mu \Bigg(&1 - \frac{\alpha^{2}}{2 n^{2}}
    - \frac{\alpha^{4}}{8 n^{4}}
    + \frac{(2 \ell - 3 n + 1) \alpha^{4}}{n^{4} (\ell + 1/2)} \nn \\
    &\quad + \frac{2 \, \tilde{a} \, m \, \alpha^{5}}{n^{3} \, \ell (\ell + 1/2)(\ell + 1)}
    + O\bigl(\alpha^{6}\bigr)\Bigg)\,,
\end{align}
while the imaginary part is expressed as
\begin{align}
\omega^{I}_{n l m} & =2 \mu r_{+}\left(m \Omega_{H}-\mu\right) \alpha^{4 l+4} \mathcal{A}_{n l} \mathcal{X}_{n l},\\
\mathcal{A}_{n l} & =\frac{2^{4 l+1}(2 l+n+1)!}{(l+n+1)^{2 l+4} n!}\left(\frac{l!}{(2 l)!(2 l+1)!}\right)^{2},\\
\mathcal{X}_{l m} & =\prod_{j=1}^{l}\left[j^{2}\left(1-\tilde{a}^{2}\right)+4 r_{+}^{2}\left(m \Omega_{H}-\mu\right)^{2}\right]\,.
\end{align}

The superradiant growth of the fields backreacts on the background geometry, causing the BH to spin down. The BH spin $\tilde{a}$ and mass $M$ evolve as
\begin{equation}
\dv{M}{t}=-\mu\sum_{n\ell m}\Gamma_{n\ell m} N_{n\ell m}\,,
\end{equation}
\begin{equation}
\dv{\tilde{a}}{t}=-m\sum_{n\ell m}\frac{\Gamma_{n\ell m} N_{n\ell m}}{G_NM^2}\,,
\end{equation}
where $\Gamma_{n\ell m}=2\omega^{I}_{n\ell m}$ is the superradiance growth rate, and $N_{\ell m}$ is the occupation number of the $\ket{n\ell m}$ state, which grows as $N_{n\ell m}\propto e^{\Gamma_{n\ell m} t}$, until the superradiance condition is no longer satisfied. At the time of saturation when an amount of spin $\Delta \tilde{a}$ has been extracted, the maximal cloud occupation number $N_{\mathrm{max}}$ in a given level can be calculated as~\cite{Arvanitaki:2014wva}
\begin{equation}
N_{\mathrm{max}} \approx 10^{76}\left(\frac{1}{m}\right)\left(\frac{\Delta \tilde{a}}{0.1}\right)\left(\frac{M}{10 \mathrm{M}_{\odot}}\right)^{2}\,.
\end{equation}

\subsection{Constraint methodology}

The existence of a rapidly growing superradiant cloud would cause a BH to spin down quickly. Therefore, the observation of a highly spinning BH can be used to exclude the existence of any ultralight boson that would trigger such an instability. A constraint can be placed on a given bosonic field if the following two conditions are met:
\begin{enumerate}
    \item A specific mode $|n\ell m\rangle$ satisfies the superradiance condition;
    \item The corresponding growth rate, $\Gamma_{n\ell m}$, is large enough for the cloud to grow into a significant size within a relevant astrophysical timescale, $\tau_{\mathrm{BH}}$ (e.g., the age of the BH).
\end{enumerate}
A simplified criterion for the second condition can be formulated by comparing the superradiance timescale, $\tau_{\mathrm{SR}} = \Gamma_{n\ell m}^{-1}$, with the BH timescale. Since the occupation number of the cloud grows exponentially, the time required to reach the maximum number of particles, $N_{\mathrm{max}}$, is proportional to $\tau_{\mathrm{SR}}\ln N_{\mathrm{max}}$. For the spin-down process to be effective, this must be shorter than the BH lifetime, leading to the condition~\footnote{It should be noted that the value of $\Delta\tilde{a}$ used to calculate $N_{\mathrm{max}}$ is typically chosen to be on the order of the observational uncertainty of the BH spin. Because the condition on $\tau_{\mathrm{SR}}$ depends only logarithmically on $N_{\mathrm{max}}$ (and thus on $\Delta\tilde{a}$), even an order-of-magnitude difference in $N_{\mathrm{max}}$ does not significantly alter the derived limits, making this timescale comparison a robust estimation method~\cite{Hoof:2024quk}.}
\begin{equation}
\tau_{\mathrm{SR}} < \frac{\tau_{\mathrm{BH}}}{\ln N_{\mathrm{max}}}\,.
\end{equation}

This constraint methodology, based on BH spin observations, is general and can be applied to various models featuring ultralight bosons. For instance, in typical axion models (potentially including self-interactions), the constraints would be placed on the axion mass $m_a$ and decay constant $f_a$~\cite{Arvanitaki:2009fg,Arvanitaki:2010sy,britoGravitationalWaveSearches2017,Stott:2018opm,Fernandez:2019qbj,Davoudiasl:2019nlo,Ng:2020ruv,Saha:2022hcd,chengConstrainsAxionlikeParticle2023,Witte:2024drg,Hoof:2024quk,Caputo:2025oap,Aswathi:2025nxa}. However, our focus in this work is to constrain the IDE models, where the relevant parameters governing the superradiance of the ultralight boson (be it DM or DE) differ due to the DE-DM coupling.

Specifically, for a given IDE model characterized by a set of fundamental parameters $\boldsymbol{\alpha}$, the superradiance predicts a maximally possible spin, $\tilde{a}^{\mathrm{crit}}$, that a BH of mass $M$ can sustain over its lifetime $\tau_{\mathrm{BH}}$. This defines an ``exclusion boundary'' or Regge trajectory in the BH parameter space, denoted as $\tilde{a}^{\mathrm{crit}}(M; \tau_{\mathrm{BH}}, \boldsymbol{\alpha})$~\cite{Hoof:2024quk}. For our Model I, the relevant parameter space is $\boldsymbol{\alpha} = (m_\chi, \beta)$, while for Model II, it simplifies into $\boldsymbol{\alpha} = (\beta)$. It is essential to note that when applying this method to our IDE models, we are currently neglecting the potential self-interactions of the superradiating field itself (analogous to the effects controlled by $f_a$ in axion models).

The physical implication remains straightforward: if an astrophysical BH is observed with a spin $\tilde{a}_{\mathrm{obs}}$ that is confidently above this theoretical maximum ($\tilde{a}_{\mathrm{obs}} > \tilde{a}^{\mathrm{crit}}$), then the specific set of model parameters $\boldsymbol{\alpha}$ that predicted that boundary is ruled out~\cite{Hoof:2024quk}. Conversely, a BH observed with a spin below the boundary ($\tilde{a}_{\mathrm{obs}} \le \tilde{a}^{\mathrm{crit}}$) is consistent with the model, as it might simply have formed with a lower initial spin~\cite{Hoof:2024quk}.

\subsection{Statistical framework}

To formalize this exclusion principle in a Bayesian context, we follow the statistical framework developed by Hoof \textit{\textit{et al.}} (2024)~\cite{Hoof:2024quk} but adapt it to our specific model and parameter space. The central idea is to compute the posterior probability for our model parameters $\boldsymbol{\alpha}$ given the observational data $D$ from a BH measurement.

The exclusion condition $\tilde{a} > \tilde{a}^\text{crit}$ can be encoded in the likelihood of observing a BH with spin $\tilde{a}$, given its mass $M$ and our model parameters $\boldsymbol{\alpha}$. This likelihood can be expressed as a Heaviside step function, $p(\tilde{a} | M, \boldsymbol{\alpha}) = \Theta(\tilde{a}^\text{crit}(M,\tau_{\mathrm{BH}},\ldots; \boldsymbol{\alpha}) - \tilde{a})$. This function is unity if the spin is at or below the critical value (allowed) and zero if it is above (excluded). We do not know the true values of the BH parameters, which we denote by the vector $\boldsymbol{\beta}_{\mathrm{BH}} = (M, \tilde{a})$. Instead, we have posterior distributions, $p(\boldsymbol{\beta}_{\mathrm{BH}}|D)$, from observations. 

To obtain the posterior for our model, $p(\boldsymbol{\alpha}|D)$, we marginalize over the unknown true BH parameters by
\begin{align}
    p(\boldsymbol{\alpha} | D) &= \int p(\boldsymbol{\alpha}, \boldsymbol{\beta}_{\mathrm{BH}} | D) \, \mathrm{d}\boldsymbol{\beta}_{\mathrm{BH}} \nonumber\\
    &= \int p(\boldsymbol{\beta}_{\mathrm{BH}} | D) \, p(\boldsymbol{\alpha} | \boldsymbol{\beta}_{\mathrm{BH}},\slashed{D}) \, \mathrm{d}\boldsymbol{\beta}_{\mathrm{BH}}\,.
\end{align}
Here we have used a slash to indicate the removal of the redundant dependence on the data D. Using the Bayes' theorem on $p(\boldsymbol{\alpha} | \boldsymbol{\beta}_{\mathrm{BH}},\slashed{D})$, we have $p(\boldsymbol{\alpha} | \boldsymbol{\beta}_{\mathrm{BH}},\slashed{D}) = p(\boldsymbol{\beta}_{\mathrm{BH}} | \boldsymbol{\alpha}) p(\boldsymbol{\alpha}) / p(\boldsymbol{\beta}_{\mathrm{BH}})$. The physical properties of the BH only depend on the model parameters through the superradiance condition, allowing us to factorize the likelihood as $p(\boldsymbol{\beta}_{\mathrm{BH}}|\boldsymbol{\alpha}) = p(\boldsymbol{\beta}_{\mathrm{BH}}) p(\tilde{a} | M,\boldsymbol{\alpha})$. Substituting this back, we find
\begin{align}
    p(\boldsymbol{\alpha} | D) &= p(\boldsymbol{\alpha}) \int \! p(\boldsymbol{\beta}_{\mathrm{BH}} | D) \, p(\tilde{a} | M, \boldsymbol{\alpha}) \, \mathrm{d}\boldsymbol{\beta}_{\mathrm{BH}}\,. \label{eq:posterior_integral}
\end{align}
In practice, observational analyses provide a set of $N$ equally weighted posterior samples, $\{\boldsymbol{\beta}_{\mathrm{BH}}^i = (M^i, \tilde{a}^i)\}_{i=1}^N$, for the BH parameters. The integral in Eq.~\eqref{eq:posterior_integral} can therefore be accurately approximated via Monte Carlo integration as
\begin{equation}
    p(\boldsymbol{\alpha} | D) \approx \frac{p(\boldsymbol{\alpha})}{N} \sum_{i=1}^{N} \Theta(\tilde{a}^\text{crit}(M^i; \boldsymbol{\alpha}) - \tilde{a}^i)\,. \label{eq:posterior_sampling}
\end{equation}

The sum effectively counts the fraction of posterior samples for the BH that are consistent with (i.e., not excluded by) the given set of model parameters $\boldsymbol{\alpha}$~\cite{Hoof:2024quk}. Since the superradiance rate $\Gamma_{\mathrm{SR}}$, and thus $\tilde{a}^\text{crit}$, depends on the properties of the ultralight boson derived from the specific IDE model (such as its effective mass, which is a function of $\boldsymbol{\alpha}$), this statistical framework allows us to translate observational data of highly spinning BHs into exclusion regions in the parameter space $\boldsymbol{\alpha}$ of the IDE model under consideration.

\section{Model I: DE as a dark fifth force mediator of the DM sector}~\label{sec:DEmediator}

We now introduce the first of our specific field-theoretic models, building upon the general IDE framework discussed in Section~\ref{sec:IDE}. In this scenario, the DE scalar field acts as a dark fifth force mediator of the DM sector, assuming both are ultralight scalar fields. This interaction alters the DM effective mass based on the background DE field, thereby modifying the superradiance potential of the DM particles themselves around spinning BHs and enabling constraints on the DE-DM coupling.

\subsection{Trilinear coupled quintessence model}

We begin by considering a coupled quintessence model where a DE scalar field, $\phi$, is coupled to an ultralight DM field~\footnote{The assumption that DM is an ultralight scalar is motivated by the naturalness of the model~\cite{Archidiacono:2022iuu}. Theoretical consistency requires that the parameters of the $\phi$-potential are at least as large as the radiative corrections induced by the interaction with DM. This leads to the condition
$$
m_{\phi}^2\gtrsim \frac{\beta}{(4\pi)^2} \frac{m_{\chi}^4}{M_\mathrm{Pl}^2}\,.
$$
Imposing $m_{\phi}\lesssim H_{0}$ yields an upper bound on the DM mass,
$$
m_{\chi} \lesssim \beta^{-1 / 4}\left(4 \pi m_{\varphi} M_\mathrm{Pl}\right)^{1 / 2} \simeq 0.02\ \mathrm{eV}\left(\frac{0.01}{\beta}\right)^{1 / 4}\left(\frac{m_{\varphi}}{H_{0}}\right)^{1 / 2}\,.
$$
Therefore, a relativistic field-theoretical description of the dark fifth force mediator without fine-tuning requires the DM to be an ultralight boson. In this mass range, the DM candidate could be an axion or axionlike particle produced via the misalignment mechanism or other nonthermal processes.}, $\chi$, via a trilinear interaction term~\cite{Farrar:2003uw,Archidiacono:2022iuu,Bottaro:2023wkd,Bottaro:2024pcb}. The Lagrangian for this model is given by
\begin{equation}
\mathcal{L}=-\frac{1}{2}\partial_{\mu}\chi\partial^\mu \chi-\frac{1}{2}m_{\chi}^2\chi^2-\frac{1}{2}\partial_{\mu}\phi\partial^{\mu}\phi-\frac{1}{2}m_{\phi}^2\phi^2-\kappa \phi \chi^2\,,
\end{equation}
where $\kappa$ is the coupling constant. This trilinear interaction term induces DM self-interactions mediated by the DE field $\phi$. This new force acts exclusively on DM and can be interpreted as a ``dark fifth force.'' In the nonrelativistic limit, the exchange of the scalar $\phi$ generates a Yukawa-type potential between DM particles,
\begin{equation}
V(r)\propto \frac{e^{-m_{\phi}r}}{r}\,,
\end{equation}
with the characteristic range set by the Compton wavelength of the scalar, $\lambda_{\phi}\sim m_{\phi}^{-1}$. 

To better analyze the interaction strength, it is convenient to re-parameterize the model. We introduce a dimensionless field $s=G_{s}^{1/2}\phi$, which allows the Lagrangian to be rewritten as
\begin{equation}
\mathcal{L}=-\frac{1}{2}(\partial \chi)^{2}-\frac{1}{2}m_{\chi}^{2}(s) \chi^{2}-\frac{1}{2G_{s}}\left[(\partial s)^{2}+m_{\phi}^{2} s^{2}\right]\,.
\end{equation}
Here, the DM mass becomes field-dependent, with an effective mass squared given by $m_{\chi}^2(s)=m_{\chi}^2(1+2s)$. The constant $G_s$ is defined as
\begin{equation}
G_{s}=\frac{\kappa^2}{m_{\chi}^4}\,,
\end{equation}
which describes the intrinsic strength of the scalar force, analogous to the Newtonian constant $G_{N}$. This allows us to define a dimensionless ratio $\beta$ to characterize the strength of this fifth force relative to gravity by
\begin{equation}
\beta=\frac{G_{s}}{4\pi G_{N}}\,.
\end{equation}
While more general interactions of the form $V_{\mathrm{int}}\sim \chi^2F(s)$, with $F(s)$ some function of the DE field, are possible, this work focuses on the simplest and illustrative case of a trilinear coupling.

Models of this type, where an ultralight scalar field (which may or may not be the DE field) mediates a long-range fifth force exclusively within the DM sector, have been extensively studied in the cosmological literature~\cite{Wetterich:1994bg,Amendola:1999er,Farrar:2003uw,Archidiacono:2022iuu,Bottaro:2023wkd,Bottaro:2024pcb}. These studies typically focus on constraining the model parameters using cosmological observations. Indeed, existing cosmological data, primarily from the Cosmic Microwave Background~\cite{Archidiacono:2022iuu} and large-scale structure surveys~\cite{Bottaro:2023wkd,Bottaro:2024pcb}, already place stringent constraints on the dimensionless coupling strength, typically finding $\beta \lesssim \mathcal{O}(0.01)$. Our work complements these cosmological probes by exploring constraints from the distinct physical regime of BH superradiance.

\subsection{Effective mass correction}

The coupling between the DE field $\phi$ and the DM field $\chi$ results in the mass of the DM particle becoming dependent on the background value of the DE field, $\bar{s}$. This gives rise to an effective mass for DM, which, for our model, is given by
\begin{equation}
    m_{\chi}^2(\bar{s})=m_{\chi}^2(1+2\bar{s}),
\end{equation}
where $m_\chi$ is the bare mass of the DM particle.\footnote{{It is crucial to emphasize that the DE mediator field $\phi$ (and consequently $s$) is characterized by an extraordinarily small bare mass, typically on the order of the present-day Hubble parameter ($m_\phi \sim H_0$). The corresponding Compton wavelength is on the cosmological scale ($\lambda_\phi \sim H_0^{-1}$). Because this macroscopic wavelength is astronomically larger than the localized scale of the BH, the field cannot resolve local spatial variations in the DM density. Therefore, the mediator field remains strictly homogeneous across the superradiant region, justifying our treatment of $\bar{s}$ as a spatially constant background value.}}

Under the hypothesis that DM is an ultralight boson, it can trigger superradiant instabilities in the vicinity of a rotating BH whose efficiency is critically dependent on the effective mass of the particle. The DE-DM coupling introduces a novel aspect: the local effective mass of a DM particle is determined by the value of the cosmological background DE field, $\bar{s}$. Thus, the global cosmological dynamics of the dark sector directly impacts a local astrophysical phenomenon. Consequently, DE-DM coupling modifies the conditions for superradiant instabilities to form around rotating BHs.

In principle, the value of the background field $\bar{s}$ evolves with cosmic time, governed by its equation of motion (see Eq~(3.2) in \cite{Archidiacono:2022iuu}). A full analysis would require solving this equation over cosmic history, assuming a fixed initial value (for a scalar mass $m_{\phi}\sim H_{0}$, one finds $\bar{s}_{\mathrm{ini}} \sim \mathcal{O}(1)$). However, since the BH sample we selected formed very recently—roughly $\mathcal{O}(10-10^2)\,\mathrm{Myr}$ ago, and the superradiance timescale is too short for significant change of $\bar{s}$, we can approximate $\bar{s}$ with its constant, present-day value $\bar{s}_{0}$ when applying superradiance constraints.


We can estimate $\bar{s}_{0}$ by making the reasonable assumption that the potential energy of the scalar field today accounts for the observed DE density, $\rho_{\mathrm{DE}}$,
\begin{equation}
    V(\bar{s}_{0})\approx \rho_{\mathrm{DE}}\approx 0.7 \times 3H_{0}^2 M_\mathrm{Pl}^2\,,
\end{equation}
where $M_\mathrm{Pl}=(8\pi G_{N})^{-1}$ is the reduced Planck mass. The potential $V(\bar{s})$ can be expressed in terms of the model parameters as
\begin{equation}
    V(\bar{s}) = \frac{1}{2}m_{\phi}^2\bar{\phi}^2 = \frac{M_\mathrm{Pl}^2 m_{\phi}^2 \bar{s}^2}{\beta}.
\end{equation}
Solving for $\bar{s}_0$ yields
\begin{equation}
    \bar{s}_{0} \approx 1.45\frac{H_{0}}{m_{\phi}}\beta^{1/2}.
\end{equation}

For the benchmark case where the mass of the DE mediator is on the order of the Hubble scale, $m_{\phi} \simeq H_{0}$, the present-day field value simplifies to $\bar{s}_{0}\simeq 1.45\beta^{1/2}$. This analytical estimate shows agreement with full numerical solutions of the cosmological evolution, as demonstrated in~\cite{Archidiacono:2022iuu}. Consequently, the effective mass correction for DM can be expressed directly in terms of the fundamental coupling strength $\beta$ by
\begin{equation}
    \Delta m_{\chi}^2 \simeq 2.9 \frac{H_{0}}{m_{\phi}} \beta^{1/2} m_{\chi}^2.\label{masscorrection}
\end{equation}
This result establishes a clear relationship: the strength of the DE-DM coupling, quantified by $\beta$, determines the magnitude of the DM effective mass correction. Since BH superradiance is highly sensitive to this mass, observations (or nonobservations) of superradiant effects can be used to place constraints on $\beta$. In the subsequent sections, we will detail the methodology for using superradiance as a probe to constrain this interaction.

\subsection{Applying superradiance constraints}

In the previous subsection, we established that the interaction between DE and DM induces an effective mass for the ultralight DM particle, which depends on the coupling strength $\beta$. The phenomenon of BH superradiance is critically sensitive to the mass of the bosonic field. This connection provides a powerful opportunity: we can leverage observations of spinning BHs to probe the DE-DM interaction itself.

\begin{figure*}[htbp]
    \centering
    \subfloat{\includegraphics[width=0.49\textwidth]{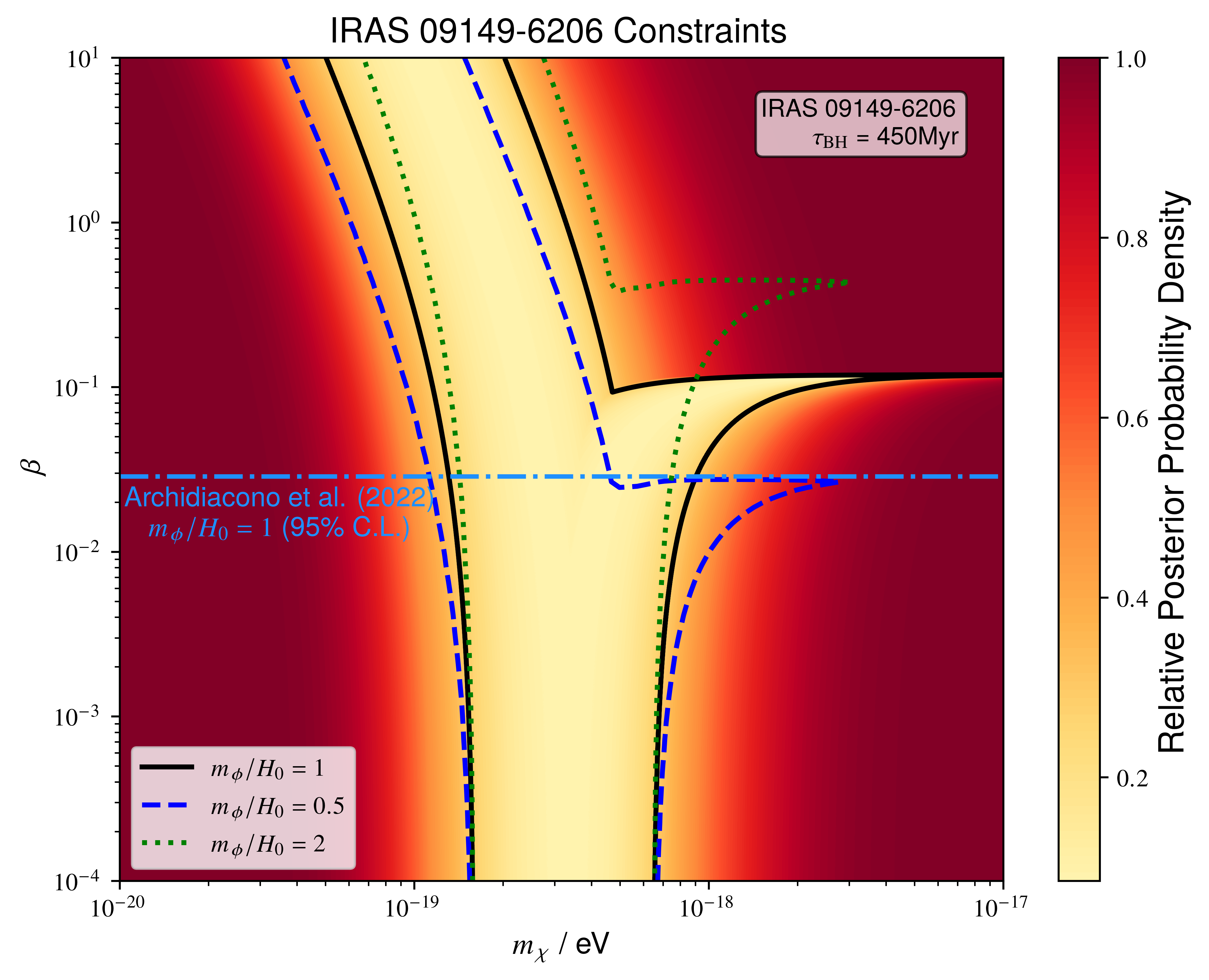}}\hfill
    \subfloat{\includegraphics[width=0.49\textwidth]{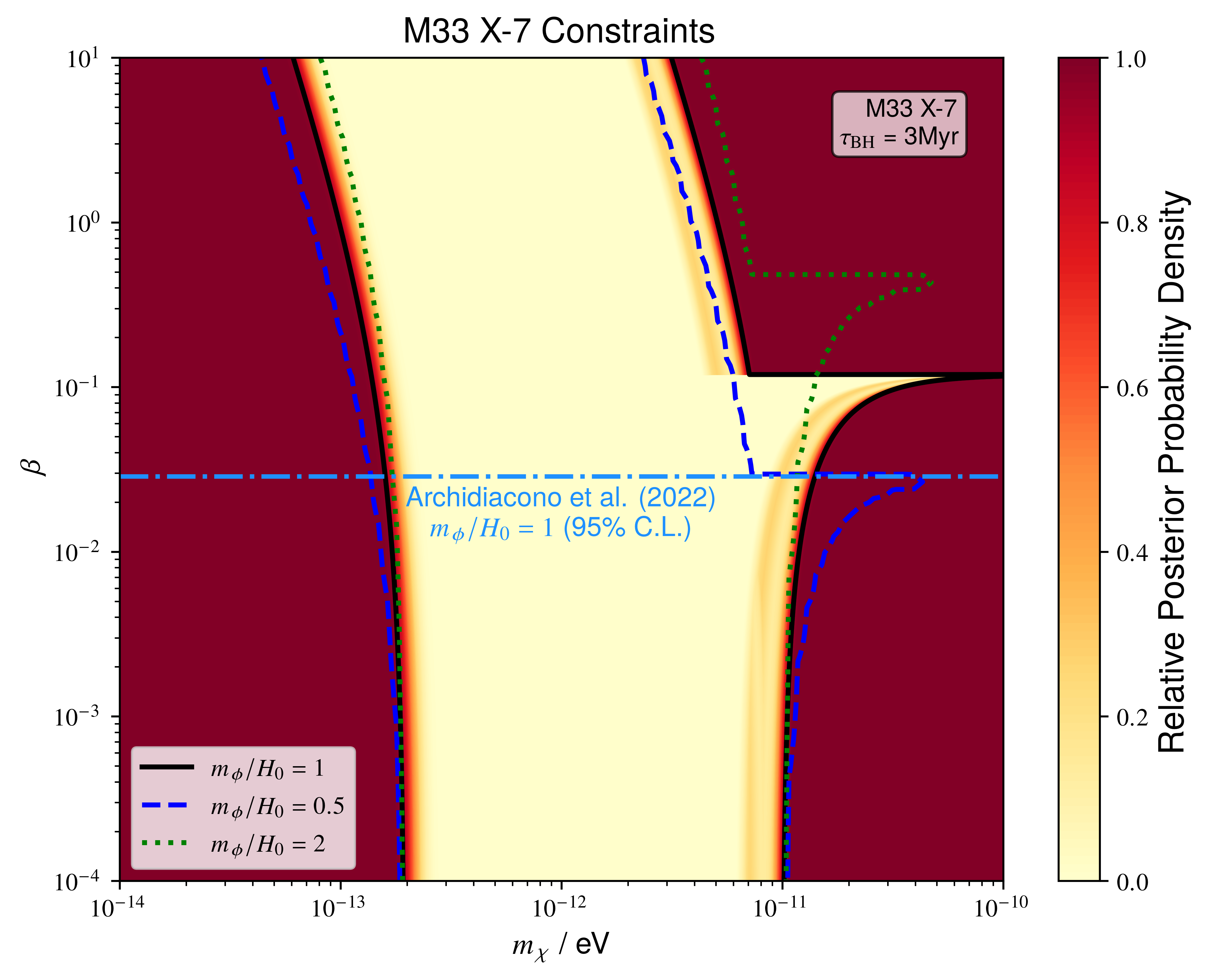}}
    \caption{Normalised posterior probability density distributions for the DM mass $m_\chi$ and the coupling strength $\beta$, derived from observations of IRAS 09149-6206 (left) and M33 X-7 (right). The filled contours represent the results for our benchmark model with the DE mediator mass set to $m_\phi/H_0=1$, and the solid black lines indicate the corresponding 95\% CRs at highest posterior density. For comparison, the dashed lines show the 95\% CRs for models with alternative mediator masses of $m_\phi/H_0=2$ and $m_\phi/H_0=0.5$. Finally, the horizontal dashed line indicates the 95\% C.L. cosmological upper limit on $\beta$ from Archidiacono et al. (2022)~\cite{Archidiacono:2022iuu}.}
    \label{fig:mchibeta_constraint}
\end{figure*}

To constrain the model parameters, we apply the statistical framework and methodology detailed in Section~\ref{sec:sr}. We utilize the BH posterior samples redistributed by Hoof \textit{et al.} (2024)~\cite{Hoof:2024quk}, specifically focusing on two systems: the stellar-mass X-ray binary BH M33 X-7~\cite{Liu:2008tk} and the supermassive black hole (SMBH) IRAS 09149-6206~\cite{GRAVITY:2020cli,Walton:2020vng,abuter2017first}. The analysis framework and associated code are based on the work presented in~\cite{Hoof:2024quk}~\footnote{The code used for the analysis is publicly available at \url{https://github.com/sebhoof/bhsr}.}. A crucial input for the superradiance constraint is the BH timescale, $\tau_{\mathrm{BH}}$, which represents the time available for the instability to develop. For accreting BHs, a conservative and physically motivated choice is the Salpeter time, $\tau_{\mathrm{Salp}} \approx 4.5 \times 10^7$ yr, which represents the characteristic timescale for a BH to significantly increase its mass or spin up via Eddington-limited accretion (assuming a canonical 10\% radiative efficiency, $
\bar{\lambda}_{\mathrm{Edd}}=0.1$). For M33 X-7, we adopt its estimated age as the timescale ($\tau_{\mathrm{BH}} = \tau_{\mathrm{age}} = 3\times 10^6 \mathrm{yr}$~\cite{Gou:2011nq}), while for the SMBH IRAS 09149-6206, we use the Eddington timescale ($\tau_{\mathrm{BH}} = \tau_{\mathrm{Edd}} = \tau_{\mathrm{Salp}}/\bar{\lambda}_{\mathrm{Edd}} = 4.5\times 10^8 \mathrm{yr}$~\cite{Hoof:2024quk}), consistent with the Salpeter time.

Our main results are presented in Figure~\ref{fig:mchibeta_constraint}, which displays the posterior probability distributions in the two-dimensional parameter space of the DM bare mass $m_\chi$ and the DE-DM coupling strength $\beta$. The constraints are derived from the IRAS 09149-6206 (left panel) and M33 X-7 (right panel) data, respectively. In our analysis, we assume log-uniform prior distributions for both $m_\chi$ and $\beta$ within the ranges shown in Figure~\ref{fig:mchibeta_constraint} and include superradiant modes up to the principal quantum number $n=8$. The solid black lines indicate the 95\% credible regions (CRs) derived from the highest posterior density for our benchmark model, where the mass of the DE mediator is set to $m_\phi = H_0$. The parameter space enclosed by the solid black lines represents the excluded region. As can be seen, for small values of $\beta$, the DE-induced correction to the DM effective mass is negligible. The constraint thus degenerates to the standard superradiance exclusion for a specific range of the DM bare mass $m_\chi$ which appears independent of $\beta$. As $\beta$ becomes larger, the effective mass correction becomes significant, causing the excluded region to shift away from this initial bare mass interval. 

To investigate the impact of the mediator mass, we also show the 95\% CRs for models with $m_\phi = 2H_0$ and $m_\phi = 0.5H_0$ as dashed lines. Since the effective mass correction is proportional to $H_0/m_\phi$, a smaller DE mass (e.g., $m_\phi = 0.5H_0$) enhances the correction term for a given $\beta$. This causes the significant shift in the exclusion region to begin at a smaller value of $\beta$. Conversely, a larger mediator mass (e.g., $m_\phi = 2H_0$) suppresses the correction, requiring a larger $\beta$ to produce the same effect, thus shifting the turn-off point to higher $\beta$ values.

Our constraints can be directly compared with those from large-scale cosmological observations. The model presented here is identical to the Coupled Dark Energy model constrained by Archidiacono \textit{et al.} (2022)~\cite{Archidiacono:2022iuu} using a combination of Cosmic Microwave Background data from Planck and Baryon Acoustic Oscillations data. Their analysis provides a $m_\chi$-independent upper bound on the coupling strength, $\beta \lesssim 0.00287$ at a 95\% confidence level (C.L.). In Figure~\ref{fig:mchibeta_constraint}, we superimpose this cosmological constraint as a horizontal line.

The constraints presented are inherently two-dimensional, providing a joint exclusion on the $(m_\chi, \beta)$ plane. To derive a specific limit on the coupling strength $\beta$ alone, prior knowledge of the DM mass $m_\chi$ is required. Fortunately, such information can be obtained from complementary cosmological and astrophysical observations. For example, studies of the Lyman-$\alpha$ forest provide robust lower bounds on the mass of ultralight scalar DM, as its quantum pressure suppresses the formation of small-scale structures. Current Lyman-$\alpha$ observations typically yield a lower bound $m_\chi \gtrsim 10^{-21} \mathrm{eV}$ for the scalar DM mass~\cite{Kobayashi:2017jcf}. By combining such independent constraints on $m_\chi$ with our superradiance results, it is possible to project the 2D exclusion region onto a one-dimensional constraint on the dark fifth force parameter $\beta$.

It is important to emphasize that the results presented here, utilizing a simple trilinear coupling model and two exemplary BH samples, serve as a demonstration of a broader methodological framework. Our primary goal is to establish that BH superradiance can be a powerful probe of DE-DM interactions where the background DE field modifies the effective mass of ultralight DM. This framework is generalizable and can be applied to more complex interaction models, and can incorporate a larger ensemble of BH observations to strengthen the constraints.

\section{Model II: DE superradiance induced by DM spike}~\label{sec:DESR}

In this section, we explore a different and novel scenario where the DE field itself, rather than the DM, is the ultralight boson that triggers BH superradiance. This is achieved through a nonminimal coupling between the DE quintessence field $\Phi$ and the DM sector. This coupling induces an effective mass for the DE scalar that is dependent on the local DM density.

\subsection{nonminimally coupled quintessence
model}

The model is defined by an action of Einstein gravity with a scalar field $\Phi$ for the DE sector,
\begin{align}
S &= \int \mathrm{d}^4 x \sqrt{-g}\left[\frac{R}{16\pi G_N}-\frac{1}{2} g_{\mu \nu} \partial^{\mu} \Phi \partial^{\nu} \Phi - V(\Phi)\right] \nn \\
  & + S_{\mathrm{DM}}\left(\Psi ; \Omega^{2}(\Phi) g_{\mu \nu}\right)\,.
\end{align}
where the DM sector couples to Einstein gravity nonminimally via this DE field $\Phi$.
Here, the conformal factor $\Omega(\Phi)$ rescales the metric experienced by DM particles $\Psi$. This structure is characteristic of certain scalar-tensor theories of gravity. The key consequence of this coupling is that the equation of motion for the DE field $\Phi$ gains a source term from the DM energy-momentum tensor, leading to an effective potential.  For pressureless DM~\footnote{Notably, unlike our previous model, this mechanism does not require DM itself to be an ultralight boson; it can be any form of cold, pressureless matter.}, the equation of motion for $\Phi$ simplifies to $\square\Phi=V'_{\mathrm{eff}}(\Phi)$, where the effective potential is $V_{\mathrm{eff}}(\Phi)=V(\Phi)+\Omega(\Phi)\rho_{\mathrm{DM}}$. Here, $\rho_{\mathrm{DM}}$ is the local energy density of DM. The details of derivation can be found in Appendix~\ref{app:effpotential}.

While models of this type are widely used in fifth-force~\cite{Uzan:2023dsk,Pitrou:2023swx,Hu:2025uqb} and cosmological studies~\cite{Wang:2025znm}, the astrophysical behavior of the scalar field $\Phi$ in the strong gravity regime near BHs has also garnered significant attention within the context of scalar-tensor theories. It has been shown that surrounding matter can induce tachyonic instabilities leading to spontaneous scalarization or, relevantly for this work, trigger superradiant instabilities by providing an effective mass to the scalar field~\cite{cardosoMatterKerrBlack2013,Cardoso:2013fwa}. This phenomenon has been explored in various astrophysical environments, such as accretion disks~\cite{Lingetti:2022psy} and DM halos~\cite{Tanaka:2025bfl}.

Our work bridges these two research areas by specifically considering the scalar field $\Phi$ as the cosmological DE and investigating its superradiant instability when triggered by the dense dark matter spikes expected to form around spinning BHs, as detailed in the next subsection. This approach places the model within both its cosmological IDE context and its astrophysical strong-gravity context, utilizing superradiance as a novel constraint mechanism derived directly from the model structure.

\subsection{DE superradiance induced by DM spike }

For our analysis, we adopt a simple low-energy effective quadratic potential for the DE field, $V(\Phi)=\frac{1}{2}\mu_{0}^2\Phi^2$, and a quartic conformal coupling function,
\begin{equation}
\Omega(\Phi)=1+\frac{1}{2}\beta \left( \frac{\Phi}{M_\mathrm{Pl}} \right)^2.
\end{equation}
Here $\beta$ is a dimensionless coupling parameter. With these choices, the equation of motion for $\Phi$ near a spinning BH~\footnote{Throughout this analysis, we assume the energy densities of the dark fields are subdominant to the central BH mass, allowing us to neglect their backreaction and adopt the vacuum Kerr solution as a fixed background metric.} becomes that of a scalar field with a position-dependent effective mass,
\begin{equation}
\square_{\mathrm{Kerr}}\Phi = \mu^2_{\mathrm{eff}}(\mathbf{r})\Phi,
\end{equation}
where the effective mass squared is given by
\begin{equation}
\mu_{\mathrm{eff}}^2(\mathbf{r})=\mu_{0}^2+\frac{\beta}{M_\mathrm{Pl}^2}\rho_{\mathrm{DM}}(\mathbf{r})\,.
\end{equation}
The vacuum mass of a quintessence field is typically of the order of the Hubble constant, $\mu_{0}\sim H_{0}\sim 10^{-33}\,\mathrm{eV}$, which is far too small to trigger superradiance around astrophysical BHs. Even in our Galactic neighborhood, where $\rho_{\mathrm{local}}\sim 0.4\,\mathrm{GeV/cm^3}$, the mass correction is negligible ($\Delta \mu_{\mathrm{eff}}^2\sim 10^{-60}\beta^{1/2}\,\mathrm{eV}^2$).

However, the immense gravitational pull of an SMBH can dramatically increase the DM density in its vicinity, forming a structure known as a DM spike, where $\rho_{\mathrm{spike}}\gg \rho_{\mathrm{local}}$~\cite{PhysRevLett.83.1719}. This extreme density enhancement provides a mechanism to significantly boost the DE effective mass. A DM halo with an initial central density profile $\rho(r) = \rho_0(r/r_0)^{-\gamma}$ will form a spike around an SMBH with a profile given by~\cite{PhysRevLett.83.1719}
\begin{equation}
\rho_{\mathrm{sp}}(r)=\rho_{R}\Theta(r-4R_{s})\left(1-\frac{4 R_{\mathrm{s}}}{r}\right)^{3}\left(\frac{R_{\mathrm{sp}}}{r}\right)^{\gamma_{\mathrm{sp}}},
\end{equation}
where $R_{\mathrm{s}}=2GM$ is the Schwarzschild radius of the SMBH; $\gamma_{\mathrm{sp}}=(9-2 \gamma) /(4-\gamma)$ which yields $2.25<\gamma_{\mathrm{sp}}<2.5$ for $0<\gamma<2$. $\rho_R$ and $R_{\mathrm{sp}}$ is the spike parameter which depend on $\gamma$ and the BH mass $M$ through the relation $R_{\mathrm{sp}}\left(\gamma, M\right)=\alpha_{\gamma} r_{0}\left(M /\left(\rho_{0} r_{0}^{3}\right)\right)^{1 /(3-\gamma)}$ and $\rho_{R}=\rho_{0}\left(R_{\mathrm{sp}} / r_{0}\right)^{-\gamma}$, where the normalization $\alpha_{\gamma}$ is numerically derived for each power-law index $\gamma$, e.g., $\alpha_{1}\approx 0.1$ and $\alpha_{2}\approx 0.02$~\cite{PhysRevLett.83.1719}. We determine the initial $\rho_0$ and $r_0$ through the mass-velocity-dispersion relationship following~\cite{Nishikawa:2017chy} (see Appendix~\ref{app:spikepara} for more details).

\begin{figure}[!h]
    \centering
    \includegraphics[width=0.45\textwidth]{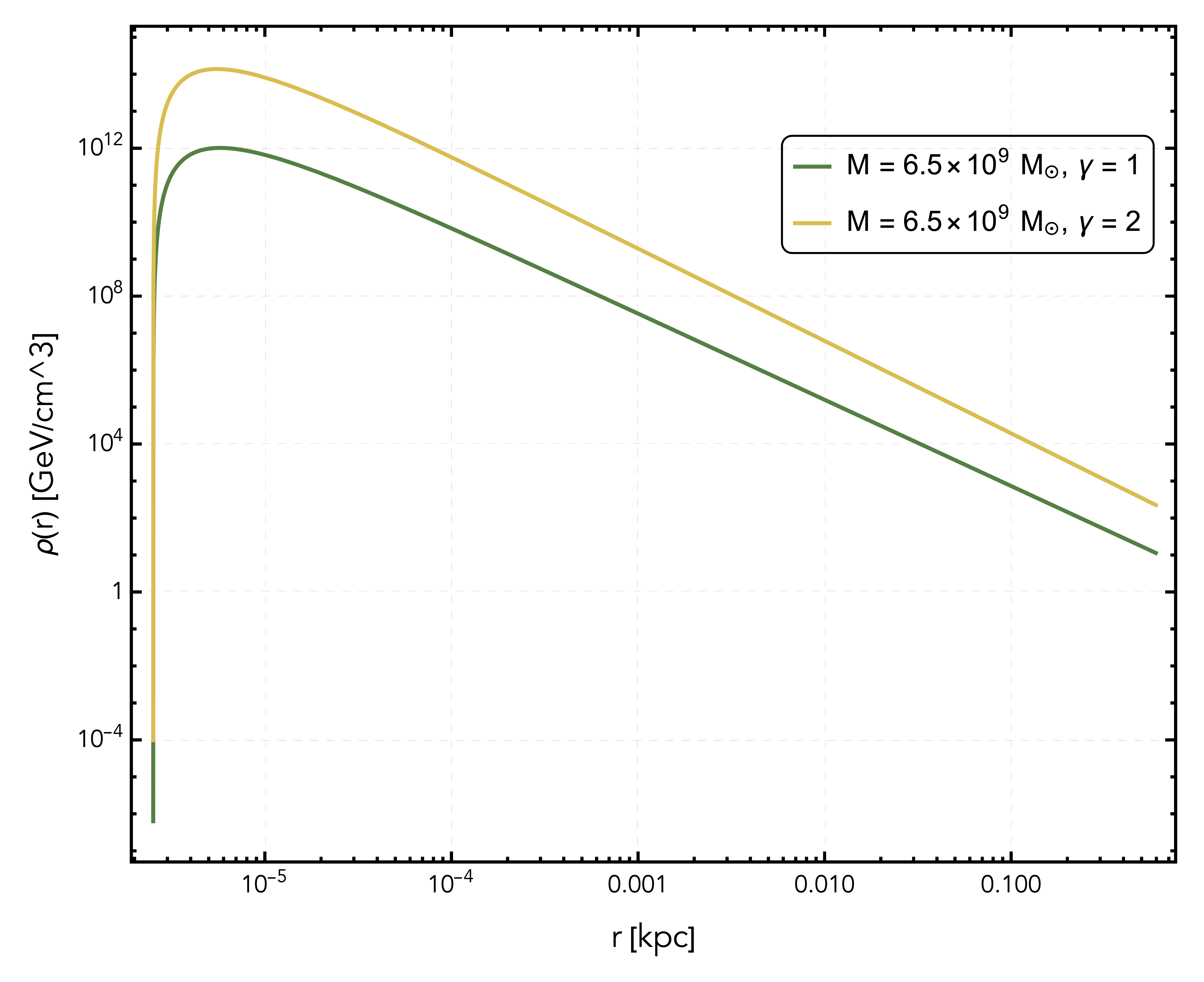}
    \caption{Dark matter density profiles around a SMBH with the mass of M87* ($M = 6.5 \times 10^9 \mathrm{M}_{\odot}$). The solid line represents the DM spike profile resulting from an initial NFW halo ($\gamma=1$, $\gamma_{\mathrm{sp}}=7/3$), while the dashed line shows the steeper spike resulting from an initial $\gamma=2$ halo ($\gamma_{\mathrm{sp}}=5/2$). The density is shown as a function of the radial coordinate $r$, covering the range from the event horizon $r_+$ to the spike radius $R_{\mathrm{sp}}$.}\label{fig:density_profile}
\end{figure}

In this work, we will consider two representative cases for the initial halo profile. The first is the standard Navarro-Frenk-White (NFW) profile~\cite{Dubinski:1991bm}, corresponding to $\gamma=1$, which results in a spike density that scales as $\rho_{\mathrm{sp}}(r) \propto r^{-7/3}$. As a second, more optimistic case for generating a strong superradiance signal, we consider a steeper initial profile with $\gamma=2$, leading to an even denser spike with $\rho_{\mathrm{sp}}(r) \propto r^{-5/2}$. Since a steeper initial profile produces a denser DM spike, a smaller value of the coupling $\beta$ is required to enhance the DE effective mass to the superradiance threshold. We will derive constraints for both the $\gamma=1$ and $\gamma=2$ scenarios. The resulting DM spike density profiles for these two cases, calculated using the mass of M87*, are illustrated in Figure~\ref{fig:density_profile}.

It is important to note that the precise profile of any DM spike is subject to significant theoretical uncertainties. A robust constraint on this model would require independent knowledge of the DM distribution. However, the primary goal of this work is to demonstrate a possible observational avenue. For a SMBH like M87*~\cite{EventHorizonTelescope:2019ggy} with a mass of $M \approx 6.5 \times 10^9 \mathrm{M}_{\odot}$ and $\gamma=1$, the DM density in the spike can be enhanced to extreme values, reaching up to $\rho_{\mathrm{max}}\sim \mathcal{O}(10^{12})\,\mathrm{GeV/cm^3}$ near the BH. This density can boost the DE effective mass to $\mu_{\mathrm{eff}}^{\mathrm{max}}\sim \mathcal{O}(10^{-24})\beta^{1/2}\,\mathrm{eV}$. For superradiance to be efficient for M87*, the boson mass must be $\mu \sim (G_NM)^{-1}\sim 10^{-20}\,\mathrm{eV}$. This implies that a relatively large coupling, $\beta \sim \mathcal{O}(10^8)$, would be required to enhance the DE mass to the necessary level. If we take $\gamma=2$, then $\beta$ would need to be about $\mathcal{O}(10^6)$.

A significant mass enhancement could trigger a powerful superradiant instability, rapidly spinning down the central BH. In this work, we use M87* as an illustrative example, with a mass of $6500.0 \pm 700.0 \times 10^6 \mathrm{M}_{\odot}$ and a spin parameter of $0.9^{+0.05}_{-0.05}$ (95\% CL)~\cite{Tamburini:2019vrf}. Thus, the observed high spin for M87* constrains the coupling parameter $\beta$~\footnote{Although M87* is thought to be rapidly spinning, its exact spin rate remains uncertain.}.

We must note a key simplification in this approach: we adopt a static, spherically symmetric DM spike profile. A fully self-consistent treatment would require modeling a dynamic, spin-dependent DM profile that co-evolves with the BH, which is beyond the scope of this work. We proceed with estimation to demonstrate the principle and potential of this novel observational avenue.

\subsection{Numerical method for spike-induced superradiance}

To find the superradiant modes, we must solve the Klein-Gordon equation for the DE field $\Phi$ in the Kerr spacetime,
\begin{equation}
    \left[\square_{\mathrm{Kerr}} - \mu^2_{\mathrm{eff}}(r)\right]\Phi(t, r, \theta, \phi) = 0,
\end{equation}
but with a position-dependent effective mass induced by the DM spike.
Here, we neglect the small vacuum mass $\mu_0$ and write the effective mass squared directly in terms of the DM spike density profile $\rho_{\mathrm{sp}}(r)$, $\mu_{\mathrm{eff}}^2(r) = \beta\rho_{\mathrm{sp}}(r)/M_\mathrm{Pl}^2$.
This position-dependent mass term acts as an effective potential for the DE field. To solve the wave equation, we use the standard separation of variables ansatz for a scalar field in Kerr spacetime,
\begin{equation}
    \Phi(t, r, \theta, \phi)=e^{-i \omega t} e^{i m \phi} R(r) S(\theta).
\end{equation}
This separates the Klein-Gordon equation into an angular equation and a radial equation. The angular part is the standard spheroidal harmonic equation,
\begin{equation}
    \frac{1}{\sin \theta} \dv{\left(\sin \theta \partial_{\theta} S(\theta)\right)}{\theta}+\left(\Lambda_{\ell m}+a^{2} \omega^{2} \cos^{2} \theta-\frac{m^{2}}{\sin^{2} \theta}\right) S(\theta)=0\,,
\end{equation}
where $\Lambda_{\ell m}$ is the separation constant. The radial part becomes a modified Teukolsky equation,
\begin{equation}
    \Delta \dv{r}\left(\Delta \dv{R(r)}{r}\right)+\left[K^{2}(r)-\left(\lambda+\mu_{\mathrm{eff}}^{2}(r) r^{2}\right) \Delta\right] R(r)=0\,,\label{radialeq}
\end{equation}
where $\Delta = r^2 - 2GMr + a^2$, $a=J/M$, $K(r)=(r^2+a^2)\omega - am$, and $\lambda=\Lambda_{\ell m}+a^{2}\omega^{2}-2am\omega$. Our goal is to find the complex eigenfrequencies $\omega = \omega_R + i\omega_I$ corresponding to the quasibound states of the field. A mode is superradiantly unstable if its growth rate $\omega_I > 0$, which occurs only when the frequency condition $0 < \omega_R < m\Omega_H$ is satisfied, where $\Omega_H$ is the angular velocity of the BH horizon.

To solve this system, we employ a hybrid numerical approach. The angular equation is standard and can be solved efficiently using the continued fraction method~\cite{Leaver:1985ax,Dolan:2007mj,Liu:2022ygf} to determine the eigenvalue $\Lambda_{\ell m}$ for a given $a\omega$. However, this method does not apply to the radial equation due to the nontrivial radial dependence of the effective mass term $\mu_{\mathrm{eff}}^2(r)$. We therefore solve the radial equation using a direct integration or ``shooting'' method~\cite{Molina:2010fb}. This involves numerically integrating the radial equation inwards from spatial infinity and outwards from the event horizon to a common matching point, imposing the appropriate boundary conditions for a quasibound state (purely outgoing waves at infinity and purely ingoing waves at the horizon). The details of the numerical method can be found in Appendix~\ref{app:numerical}. We then search for the complex frequency $\omega$ that allows for a smooth matching of the solution and its derivative at the matching point. For this analysis, we focus on the fastest-growing fundamental mode, $\ell=m=1$.

\begin{figure}[h]
    \centering
    \includegraphics[width=0.49\textwidth]{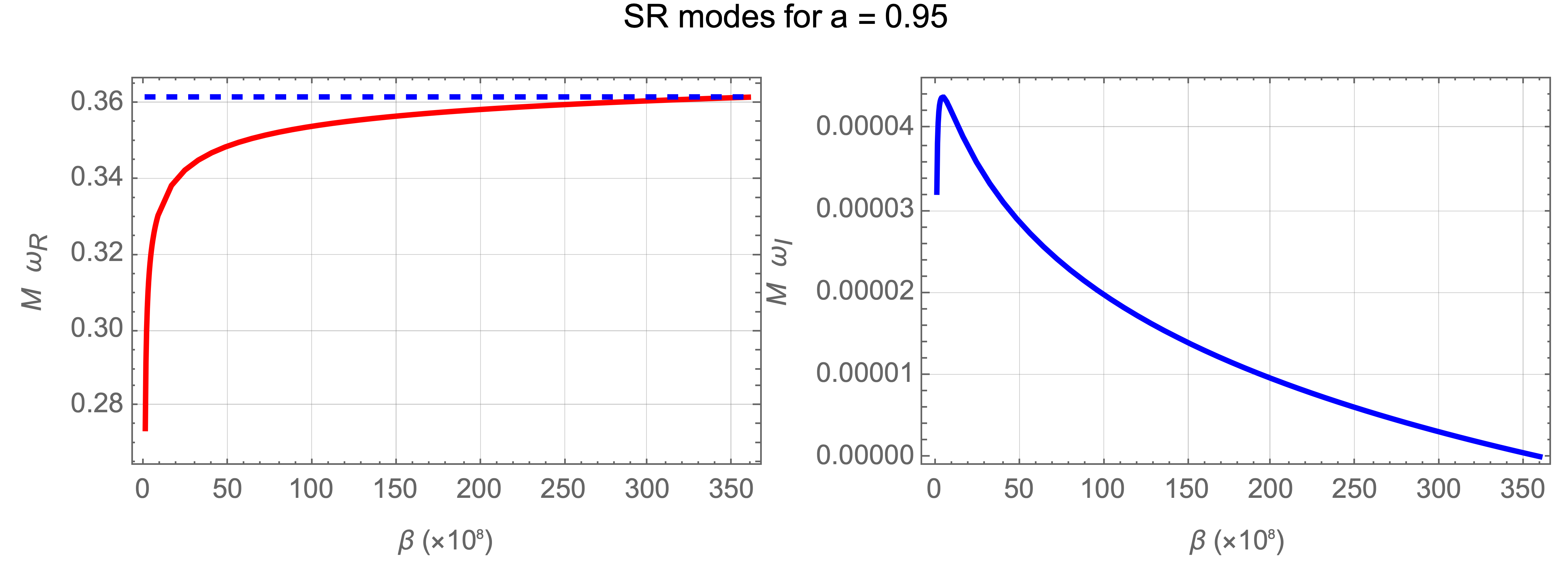}\\
    \includegraphics[width=0.49\textwidth]{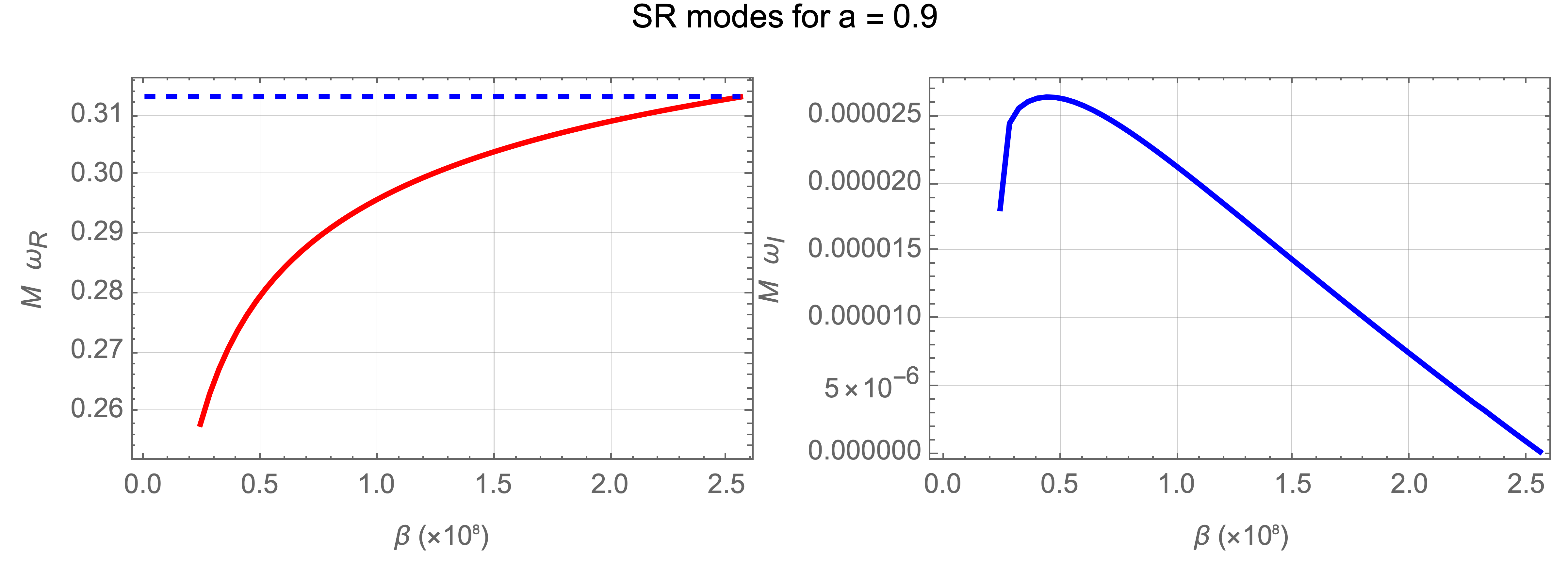}\\
    \includegraphics[width=0.49\textwidth]{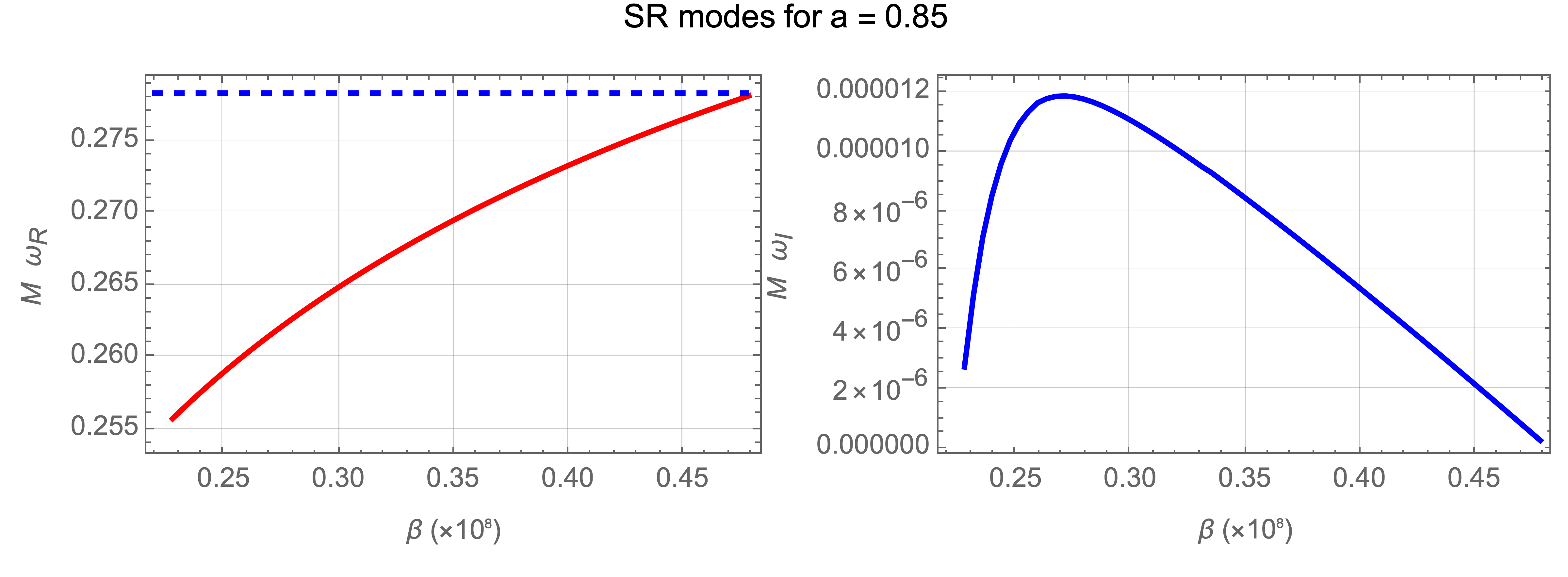}
    \caption{The real part $M\omega_R$ and imaginary part $M\omega_I$ of the frequency for the fundamental superradiant mode ($\ell=m=1$) and the $\gamma=1$ case. The frequencies are plotted as a function of the coupling parameter $\beta$ for three representative values of the BH spin: $\tilde{a}=0.95$ (top), $\tilde{a}=0.9$ (middle), and $\tilde{a}=0.85$ (bottom). The blue dashed line marks the superradiance threshold $\omega_R=m\Omega_H$. A superradiant instability exists in the region where $\omega_I > 0$.}
    \label{fig:sr_modes}
\end{figure}

We apply the numerical method described above to solve for the complex eigenfrequencies, $\omega = \omega_R + i\omega_I$, of the DE field. For a given BH mass and spin, we can determine the instability growth rate, provided by the imaginary part of the frequency $\omega_I$, as a function of the DE-DM coupling parameter $\beta$. Here we present an illustrative result for the $\gamma=1$ case and with benchmark spin values in Figure~\ref{fig:sr_modes}. The plots show the real part ($\omega_R$) and imaginary part ($\omega_I$) of the frequency for the fundamental mode, presented in units of $M^{-1}$. Each panel corresponds to a different value of the dimensionless BH spin parameter, $\tilde{a}$. As shown, for each spin value, an instability (where $\omega_I > 0$) is triggered only when the coupling $\beta$ exceeds a certain threshold. The growth rate then increases to a maximum before decreasing and eventually vanishing as $\beta$ becomes very large, which pushes $\omega_R$ outside the superradiance regime ($ \omega_R > m\Omega_H $).

The physical mechanism at play is analogous to the classic ``BH bomb'' scenario, where a massive field is trapped by a reflecting mirror placed around a rotating BH~\cite{cardosoMatterKerrBlack2013}. In our model, the dense DM spike creates a gravitational potential well that acts as the ``mirror'', trapping the DE field modes\footnote{{While a nonvanishing interior density (e.g., a ``corona'' configuration) can quench superradiant instabilities in the plasma-driven models~\cite{Dima:2020rzg,Wang:2022hra}, this does not apply here. Standard cold dark matter is strictly collisionless; particles crossing the capture radius inevitably fall into the black hole without collisional replenishment. This phase-space depletion guarantees a strictly empty cavity, preserving a robust instability.}}. The coupling strength $\beta$ controls the ``reflectivity'' of this mirror. A critical value of $\beta$ is required to create a potential barrier high enough to effectively trap the modes and trigger the instability. As $\beta$ increases further, the trapping becomes more efficient and the instability growth rate $\omega_I$ increases. However, a larger $\beta$ also increases the effective mass and thus the real frequency $\omega_R$. The instability is quenched once $\omega_R$ becomes too large and violates the superradiance condition $\omega_R < m\Omega_H$.

\subsection{Applying superradiance constraints}

We now apply the statistical framework developed in Section~\ref{sec:sr} to constrain the DE-DM coupling parameter, $\beta$, using observations of the SMBH M87*. We adopt the mass and spin values for M87* from recent observations, which indicate a high spin of $0.9^{+0.05}_{-0.05}$ (95\% CL)~\cite{Tamburini:2019vrf}. The methodology is analogous to the one previously described, but it is adapted for the specifics of the current model. A key distinction is that our parameter space is now one-dimensional, consisting solely of the coupling constant $\beta$, as the mass of the superradiating boson (the DE field) is primarily determined by $\beta$ and the local DM density.

Since full posterior distribution samples for the M87* parameters are not publicly available, we generate a mock posterior sample set. We achieve this by drawing a large number of $(M, \tilde{a})$ pairs from independent Gaussian distributions, using the reported means and standard deviations for the mass and spin. For each value of $\beta$ that we wish to test, and for each sampled $(M, \tilde{a})$ pair from our mock dataset, we perform the following procedure:
\begin{enumerate}
    \item Using the numerical method developed in the previous subsection, we solve for the complex frequency $\omega$ of the fundamental superradiant mode ($\ell=m=1$). This gives us the instability growth rate, $\Gamma_{\mathrm{SR}} = 2\omega_I$, for the given set of parameters $(M, \tilde{a}, \beta)$.
    \item We then check if this growth rate is fast enough to spin down the BH within its lifetime. We compare the superradiance timescale, $\tau_{\mathrm{SR}} = 1/\Gamma_{\mathrm{SR}}$, against the relevant astrophysical timescale for M87*, which we take to be $\tau_{\mathrm{BH}} \approx 10^9$ years following~\cite{Davoudiasl:2019nlo}.
\end{enumerate}

A specific value of the coupling parameter $\beta$ is considered excluded at a 95\% confidence level if, for more than 95\% of the sampled $(M, \tilde{a})$ pairs, the resulting superradiance timescale is shorter than the BH  lifetime ($\tau_{\mathrm{SR}} < \tau_{\mathrm{BH}}$). This statistical approach allows us to properly account for the observational uncertainties in the BH properties when deriving our constraints on the fundamental coupling constant of the IDE model.

\begin{figure}[htbp]
    \centering
    \includegraphics[width=0.49\textwidth]{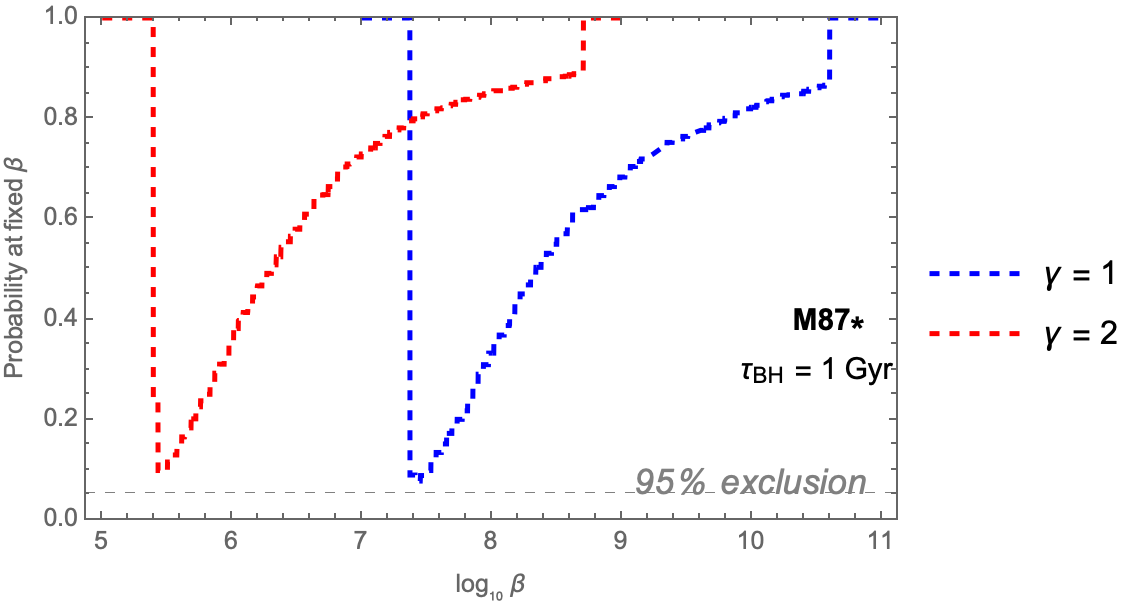}
    \caption{The normalised posterior probability density distribution for the coupling parameter $\beta$, derived from M87* observations. A log-uniform prior was assumed for $\beta$. The shaded region represents the parameter space excluded at 95\% confidence.}
    \label{fig:beta_constraint_m87}
\end{figure}

Applying the statistical framework described above to our mock M87* BH sample, we derive constraints on the DE-DM coupling parameter, $\beta$, for two distinct scenarios corresponding to the initial halo profile index: $\gamma=1$ (NFW) and a steeper profile with $\gamma=2$. For this analysis, we generated 1000 mock posterior samples for each scenario. We analyze these two cases separately, assuming a log-uniform prior for $\beta$ within the ranges of $10^7-10^{11}$ for the $\gamma=1$ case and $10^5-10^9$ for the $\gamma=2$ case.

Our main results are the posterior probability distributions for $\beta$, presented in Figure~\ref{fig:beta_constraint_m87}. The shape of the posterior effectively quantifies how each value of $\beta$ is constrained by the observation of a high spin for M87*. For values of $\beta$ below a critical threshold, $\beta_c$ (around $10^{7.4}$ for $\gamma=1$ and $10^{5.4}$ for $\gamma=2$ respectively), the model is consistent with the data, as the coupling is too weak to induce a significant superradiant instability for the majority of the BH posterior samples. As $\beta$ approaches this critical value, the superradiance rate becomes efficient for most of the high-spin samples, causing the posterior probability to drop sharply.

However, as $\beta$ increases further beyond this point, the probability begins to rise again. This is because a very large coupling results in the real part of the frequency, $\omega_R$, being pushed outside the superradiant window ($\omega_R > m\Omega_H$) for a growing fraction of the BH samples (particularly those with lower spins in the posterior).

From the posterior distributions, we find that the data are most sensitive to values of $\beta$ near the probability minima. These points of maximum sensitivity occur at $\beta \sim 10^{7.4}$ for the $\gamma=1$ (NFW) scenario and $\beta \sim 10^{5.4}$ for the steeper $\gamma=2$ scenario. We note that this specific nonminimally coupled model (with our chosen $V(\Phi)$ and $\Omega(\Phi)$) has not yet been constrained using large-scale cosmological data. A dedicated cosmological analysis is left for future exploration, making this superradiance sensitivity the first probe of this particular parameter space. Although this does not represent a strict exclusion, a limitation stemming from the constraining power of the current BH data sample, this result clearly demonstrates the powerful potential of using astrophysical observations of BHs to probe the parameter space of these fundamental interactions.

\section{Conclusions and discussions}~\label{sec:conclusion}

In this work, we have explored a novel and powerful astrophysical avenue for probing the nature of the dark sector: using BH superradiance to constrain IDE models. While traditional constraints on IDE models rely on their influence on the large-scale structure and cosmic expansion history, our approach leverages the impact of dark sector interactions on the local, small-scale physics around spinning BHs. We have demonstrated the viability of this method through the detailed analysis of two distinct, physically motivated field-theoretic models.

Our central idea is that a coupling between DE and DM can alter the effective mass of an ultralight bosonic field, which in turn significantly affects its ability to trigger superradiant instabilities. The observation of highly spinning BHs, which would have otherwise been spun down by such instabilities, can therefore be used to place constraints on the fundamental parameters of these interaction models.

In our first model, we considered a scenario where the DE scalar field acts as a mediator for a dark fifth force in the DM sector. Here, the DE-DM coupling strength, parameterized by a dimensionless parameter $\beta$, modifies the effective mass of the ultralight DM field. By applying a Bayesian statistical framework to observational data from a stellar-mass BH M33 X-7 and a SMBH IRAS 09149-6206, we have shown that superradiance can place stringent, joint constraints on the two-dimensional parameter space $(m_\chi, \beta)$. We noted that these constraints are complementary to those from cosmological observations, such as the Lyman-$\alpha$ forest, which can provide independent limits on $m_\chi$ and help break the degeneracy.

In our second model, we proposed a novel mechanism where the DE field \textit{itself} becomes the superradiating particle. In its vacuum state, the DE field is far too light to trigger superradiance around astrophysical BHs. However, we demonstrated that a nonminimal coupling to DM can dramatically enhance its effective mass in regions of high DM density. Specifically, the presence of a dense DM spike around an SMBH can increase the effective mass of the DE field to a level where superradiance becomes efficient. Using M87* as an illustrative example, we showed that the observation of its high spin places a direct upper limit on the coupling strength, excluding $\beta \sim 10^{7.4}$ for a standard NFW halo profile and $\beta \sim 10^{5.4}$ for a steeper halo profile.

For this second model, it is important to discuss the potential impact of nonlinear backreaction on the environmental DM. As the DE field grows, it exerts a gradient force that could perturb the DM density profile. Such backreaction effects are notoriously severe in the analogous context of standard plasma-driven superradiance, where nonlinearities rapidly quench the instability through the phenomenon of relativistic transparency---as the plasma is accelerated to relativistic velocities, the plasma frequency (the effective mass of photons) is suppressed by the Lorentz factor $\gamma$~\cite{Cardoso:2020nst,Blas:2020kaa,Cannizzaro:2023ltu}. Remarkably, our scalar-tensor model is inherently protected from this most detrimental quenching mechanism~\cite{Lingetti:2022psy}. The superradiant scalar field acquires its effective mass from the trace of the DM energy-momentum tensor, $T \simeq -\rho_0$. Because the rest-mass density $\rho_0$ is a Lorentz scalar, the effective mass remains strictly unsuppressed by the $\gamma$ factor, thereby entirely evading relativistic transparency. Combined with the collisionless nature of DM, which precludes the collective hydrodynamic blow-out typical of collisional plasmas, the instability in our scenario is theoretically substantially more robust. Although determining the full nonlinear dynamics requires numerical simulations, the distinct physical nature of this mechanism justifies our current static estimation as a meaningful proof of principle.

In conclusion, this work establishes BH superradiance as a potent and independent probe of the fundamental physics governing the dark sector. We have developed a general framework to connect astrophysical BH spin measurements to the parameters of field-theoretic IDE models and have provided the first such constraints for two distinct interaction scenarios.

This research opens several avenues for future investigation. The constraints are currently limited by the number of well-measured BH systems and by uncertainties in astrophysical modeling, particularly the precise density profiles of DM spikes. Future observations of BH populations from gravitational wave detectors and electromagnetic surveys will significantly expand the available data. Furthermore, our framework can be extended to a wider variety of IDE models, including more complex interaction forms and different coupling mechanisms. A combined analysis, integrating these novel astrophysical constraints with traditional cosmological data, promises to deepen our understanding of the enigmatic dark Universe significantly.

	\begin{acknowledgments}
		We thank Zhuan Ning, Zi-Yan Yuwen, and Cheng-Jun Fang for insightful discussions and comments. This work is supported by the National Key Research and Development Program of China Grants No. 2021YFC2203004, No. 2021YFA0718304, and No. 2020YFC2201501; the National Natural Science Foundation of China Grants No. 12422502, No. 12547110, No. 12588101, No. 12235019, and No. 12447101; and the China Manned Space Program Grant No. CMS-CSST-2025-A01. 
	
	\end{acknowledgments}

\section*{DATA AVAILABILITY}
The data that support the findings of this article are openly available~\cite{zenodo_data}.

\appendix

\section{Derivation of the effective energy transfer rate $Q$ from field theory}
\label{app:Q_derivation}

In this Appendix, we provide the detailed derivation of the continuity equations and the effective energy transfer rate $Q$ from the field-theoretic Lagrangian presented in Section~\ref{subsec:fieldtheoretic} and also in Section~\ref{sec:DESR}.

Similar to the assumption carried out in $\Lambda$CDM, we consider a flat, homogeneous, and isotropic Universe characterized by the Friedmann-Lemaître-Roberston-Walker (FLRW) metric,
\begin{equation}
\mathrm{d} s^2 = -\mathrm{d} t^{2}+a^{2}(t) \, \mathrm{d}\mathbf{x}^{2},
\end{equation}
where $a$ is the time-dependent scale factor. An essential ingredient for studying cosmic evolution is the Einstein field equations,
\begin{equation}
G_{\mu\nu}=8\pi G_N \, T_{\mu\nu}\,,
\end{equation}
where the stress-energy tensor $T_{\mu\nu}$ contains contributions from the $\phi$ and $\chi$ dark sectors as well as the Standard Model particles,
\begin{equation}
T_{\mu\nu}=T_{\mu\nu}^{(\phi)}+T_{\mu\nu}^{(\chi)}+T_{\mu\nu}^{\mathrm{M}}\,.
\end{equation}
When an interaction term $W(\phi, \chi)$ is present, there is an ambiguity in how to define the individual stress-energy tensors for $\phi$ and $\chi$, as only their sum is uniquely determined and conserved. A common way to partition the interaction potential is
\begin{align}
	T_{\mu\nu}^{(\phi)}&=\partial_{\mu} \phi \partial_{\nu} \phi-g_{\mu \nu}\left[\frac{1}{2} \partial^{\alpha} \phi \partial_{\alpha} \phi+V\left(\phi\right)+xW(\phi,\chi)\right], \\
	T_{\mu\nu}^{(\chi)}&=\partial_{\mu} \chi \partial_{\nu} \chi-g_{\mu \nu}\left[\frac{1}{2} \partial^{\alpha} \chi \partial_{\alpha} \chi+U\left(\chi\right)+(1-x)W(\phi,\chi)\right]\,,
\end{align}
where $x\in[0,1]$ is a real number chosen so that the sum of the energy densities of $\phi$ and $\chi$ equals the total dark sector energy density. The energy densities and pressures for the fields in a homogeneous and isotropic Universe are then read from the $00$ and $ii$ components of these tensors, respectively,
\begin{align}
\rho_{\phi} &=\frac{1}{2} \dot{\phi}^{2}+V(\phi)+xW(\phi, \chi)\,, \nn \\
P_{\phi} &=\frac{1}{2} \dot{\phi}^{2}-V(\phi)-xW(\phi, \chi) \nn \\
\rho_{\chi} &=\frac{1}{2} \dot{\chi}^{2}+U(\chi)+(1-x)W(\phi, \chi)\,, \nn \\
P_{\chi} &=\frac{1}{2} \dot{\chi}^{2}-U(\chi)-(1-x)W(\phi, \chi)\,.
\end{align}
The remaining contribution from the Standard Model particles can be modeled by a ideal fluid with the energy density $\rho_{\mathrm{M}}$, pressure $P_{\mathrm{M}}$ and the four velocity $u_{\mu}$ in a form of $T_{\mu \nu}^{M}=\rho_{M} u_{\mu} u_{\nu}+P_{M}\left(u_{\mu} u_{\nu}+g_{\mu \nu}\right)$. With these definitions, the total energy density of the dark sector is simply $\rho_{D}=\rho_{\phi}+\rho_{\chi}$. The $00$ component of the Einstein equation gives us the Friedmann equation relating the Hubble parameter $H = \dot{a}/a$ so that
\begin{equation}
3 H^{2} =8\pi G_N(\rho_{D}+\rho_{\mathrm{M}})\,.
\end{equation}

The Klein-Gordon equations for the fields are
\begin{align}
	\ddot{\phi}+3 H \dot{\phi}+V_{,\phi}+W_{,\phi}&=0\,, \\
	\ddot{\chi}+3 H \dot{\chi}+U_{,\chi}+W_{,\chi}&=0\,,
\end{align}
where a subscript $,i$ denotes a partial derivative with respect to the field $i$. Taking the time derivative of the energy densities and using the equations of motion, we can express them in the form of fluid continuity equations,
\begin{align}
&\dot{\rho}_{\phi}+3 H\left(\rho_{\phi}+P_{\phi}\right) \nn \\
&= \dot{\phi}\bigl(\ddot{\phi} + 3H\dot{\phi} + V_{,\phi}\dot{\phi} + W_{,\phi}\dot{\phi}\bigr)
    -(1-x)W_{,\phi}\dot{\phi}
    + xW_{,\chi}\dot{\chi} \nn \\
&= -(1-x)W_{,\phi}\dot{\phi}+xW_{,\chi}\dot{\chi} \equiv Q_\phi, \nn \\
&\dot{\rho}_{\chi}+3 H\left(\rho_{\chi}+P_{\chi}\right) \nn \\
&= \dot{\chi}\bigl(\ddot{\chi} + 3H\dot{\chi} + U_{,\chi}\dot{\chi} + W_{,\chi}\dot{\chi}\bigr)
    - xW_{,\chi}\dot{\chi}
    + (1-x)W_{,\phi}\dot{\phi} \nn \\
&= xW_{,\chi}\dot{\chi}+(1-x)W_{,\phi}\dot{\phi} \equiv Q_\chi\,.
\end{align}
This derivation shows that with this particular choice for splitting the stress-energy tensor, we recover $Q_\chi = -Q_\phi$, consistent with the phenomenological model in Eq.~\eqref{fluid}. When we set $x=0$, we recover the result in the main text.

However, this relation is not general. If we were to adopt a different convention, the form of $Q$ would change. For example, if we instead defined~\cite{Aboubrahim:2024spa}
\begin{align}
	T_{\mu\nu}^{(\phi)}
&=\partial_{\mu} \phi \, \partial_{\nu} \phi
    - g_{\mu \nu}\left[\frac{1}{2} \partial^{\alpha} \phi \, \partial_{\alpha} \phi
    + V\left(\phi\right)+W(\phi,\chi)\right],\\
	T_{\mu\nu}^{(\chi)}
&=\partial_{\mu} \chi \, \partial_{\nu} \chi
    - g_{\mu \nu}\left[\frac{1}{2} \partial^{\alpha} \chi \, \partial_{\alpha} \chi
    + V\left(\chi\right)+W(\phi,\chi)\right]\,,
\end{align}
then, following a similar procedure, the interaction terms would become $Q_\phi=W_{,\chi}\dot{\chi}$ and $Q_\chi=W_{,\phi}\dot{\phi}$. In this case, $Q_\phi + Q_\chi = \dot{W} \neq 0$, and the simple picture of direct energy transfer is lost~\footnote{The total energy density of the dark sector, defined as $\rho_{D}=\rho_\phi+\rho_\chi-W(\phi,\chi)$ is still conserved in this case.}. 

It is instructive to contrast this with Model II, where the interaction between the DE field ($\Phi$) and DM ($\Psi$) arises indirectly through a nonminimal coupling, specifically via the action term $S_{\mathrm{DM}}(\Psi; \Omega^{2}(\Phi) g_{\mu \nu})$ in the Einstein frame. In this scenario, there is no direct interaction potential $W(\Phi,\Psi)$ added to the Lagrangian densities of the fields. Consequently, the stress-energy tensors for the DE field, $T_{\mu\nu}^{(\Phi)}$, and the DM sector, $T_{\mu\nu}^{(\mathrm{DM})}$, are unambiguously defined based solely on their respective kinetic terms and self-interaction potentials ($V(\Phi)$ for DE, and the implicit potential within $S_{\mathrm{DM}}$).

Despite the absence of a direct potential coupling, an effective energy exchange still occurs due to the nonminimal coupling factor $\Omega(\Phi)$. Varying the total action leads to modified conservation equations (continuity equations). For pressureless DM, these take the forms of
\begin{align}
\dot{\rho}_{\phi}+3 H\left(\rho_{\phi}+P_{\phi}\right)=-Q\,, \\
\dot{\rho}_{\mathrm{dm}}+3 H\rho_{\mathrm{dm}}=Q\,,
\end{align}
where $\rho_{\Phi}$ and $P_{\Phi}$ are the standard energy density and pressure derived from the minimally coupled Lagrangian for $\Phi$, and $\rho_{\mathrm{dm}}$ is the energy density of the conformally coupled DM. The effective energy transfer rate $Q$ is found to be
\begin{equation}
	Q=\frac{\Omega'(\Phi)}{\Omega(\Phi)}\dot{\Phi}\rho_{\mathrm{dm}}\,.
\end{equation}
Thus, even in the case of indirect coupling via gravity, an effective energy transfer rate $Q$ can be derived, relating the microscopic model to the phenomenological fluid description, but without the definition ambiguity associated with direct potential interactions.

\section{Effective potential in the nonminimally coupled quintessence model}
\label{app:effpotential}

This Appendix provides details on the derivation of the effective potential for the DE quintessence field $\Phi$ in Model II, where it couples to the DM sector represented by $\Psi$ with conformal factor $\Omega(\Phi)$.

We start with the action in the Einstein frame,
\begin{align}
S &=\int \mathrm{d}^4 x \sqrt{-g^{E}}\left[\frac{R^{E}}{16\pi G}-\frac{1}{2} g_{\mu \nu}^{E} \partial^{\mu} \Phi \partial^{\nu} \Phi-{V(\Phi)}\right] \nn \\
  &+S\left(\Psi ; \Omega^{2}(\Phi) g_{\mu \nu}^{E}\right)\,.
\end{align}
Here, quantities with the superscript `E' are defined in the Einstein frame. The equation of motion is
\begin{equation}
\begin{aligned}
G_{\mu \nu}^{E} & =8 \pi G\left(T_{\mu \nu}^{E}+\partial_{\mu} \Phi \partial_{\nu} \Phi-\frac{g_{\mu \nu}^{E}}{2}(\partial \Phi)^{2}-{g_{\mu \nu}^{E}} V(\Phi)\right) \\
\square^{E} \Phi & =-\frac{\Omega^{\prime}(\Phi)}{\Omega(\Phi)} T^{E}+{V^{\prime}(\Phi)}\,,
\end{aligned}
\end{equation}
where $T^E_{\mu\nu}$ is the stress-energy tensor of DM in the Einstein frame and $T^E$ is its trace. 

The stress-energy tensor for DM in the Einstein frame is related to the one in the Jordan frame by a conformal transformation,
\begin{equation}
T_{\mu \nu}^{E}=\Omega^{2}(\Phi) T^J_{\mu \nu}\,.
\end{equation}
where $T^J_{\mu \nu}$ is the energy-momentum tensor defined in the Jordan frame where $\Psi$ is minimally coupled with gravity,  
\begin{equation}
{T}^J_{\mu \nu}=\frac{-2}{\sqrt{-{g}^J}} \frac{\partial}{\partial {g}^{J,\mu \nu}}\left(\sqrt{-{g}^J}\, \mathcal{L}_{\mathrm{dm}}\left[{g}^J_{\mu \nu}, \Psi\right]\right)\,.
\end{equation}
While $T^J_{\mu \nu}$ is covariantly conserved in the Jordan frame, i.e.,$\nabla^J_{\mu}T^{J,\mu}_{\nu}=0$, the Einstein frame tensor is generally not conserved,  
\begin{equation}
\nabla^E_{\mu} T_{\nu}^{E,\mu}=T^E \Omega^{-1} \nabla_{\nu} \Omega\,.
\end{equation}
Assuming DM behaves as a perfect fluid in the Einstein frame, $T_{\nu}^{E,\mu}=\mathrm{diag}(-\rho,p,p,p)$ with constant equation of state parameter $w$ defined by $p=w\rho$, the $\nu=0$ component reads as
\begin{equation}
\nabla_{t}\left(\Omega^{3 w-1} \rho\right)=0\,.
\end{equation}
This leads to the definition of a covariantly conserved density:
\begin{equation}
\hat{\rho}=\Omega^{3w-1}\rho\,,
\end{equation}
which is also $\Phi$-independent from $0=\nabla_{t}\hat{\rho}=\hat{\rho}'\nabla_{t}\Phi$.

The trace of the DM stress-energy tensor can then be expressed in terms of this conserved density,
\begin{equation}
T^E=(3w-1)\rho=(3w-1)\Omega^{1-3w}\hat{\rho}.
\end{equation}
Substituting this trace into the equation of motion for $\Phi$ yields
\begin{equation}
\square^{E} \Phi =(1-3w)\Omega^{\prime}\Omega^{-3w}\hat{\rho}+{V^{\prime}(\Phi)}\,.
\end{equation}
This allows us to identify an effective potential driving the dynamics of $\Phi$, 
\begin{equation}
V_{\mathrm{eff}}(\Phi)=V(\Phi)+\Omega^{1-3w}(\Phi)\hat{\rho}\,.
\end{equation}
For pressureless dark matter ($w=0$), this simplifies to $V_{\mathrm{eff}}(\Phi)=V(\Phi)+\Omega(\Phi)\hat{\rho}_{\mathrm{dm}}$ as shown in Section~\ref{sec:DESR}.

\section{Relating DM halo parameters to supermassive-black-hole mass}\label{app:spikepara}

When a BH surrounded by $\rho(r)=\rho_0(r/r_0)^{-\gamma}$ DM profile grows adiabatically, it gravitationally concentrates the surrounding DM around it to a cuspy profile usually referred to as a ``DM spike.'' The DM density spike can be modelled generally as~\cite{PhysRevLett.83.1719}
\begin{equation}
\rho_{\mathrm{sp}}(r)=\rho_{R}\Theta(r-4R_{s})\left(1-\frac{4 R_{\mathrm{s}}}{r}\right)^{3}\left(\frac{R_{\mathrm{sp}}}{r}\right)^{\gamma_{\mathrm{sp}}}.
\end{equation}
This Appendix details the procedure, following~\cite{Nishikawa:2017chy}, used to estimate the DM profile halo parameter $\rho_0$ and $r_0$ associated with a given SMBH mass
$M$. These halo parameters are necessary to calculate the structure of the DM spike, specifically $\rho_R$ and $R_{\mathrm{sp}}$, as described in Section~\ref{sec:DESR}. The method assumes the DM halo follows an NFW profile outside the region influenced by the SMBH, extending to the virial radius $r_{\mathrm{vir}}$.

First, we utilize the empirical $M$-$\sigma$ relationship, which connects the SMBH mass $M$ to the one-dimensional velocity dispersion $\sigma$ of the host bulge or halo~\cite{Gebhardt:2000fk,Ferrarese:2000se} via
\begin{equation}
\log_{10}\left(M / \mathrm{M}_{\odot}\right)=a+b \log_{10}\left(\sigma / 200 \mathrm{~km} \mathrm{~s}^{-1}\right)
\end{equation}
with empirically determined parameters $a = 8.12$ and $b = 4.24$.
Second, for an NFW halo, the velocity dispersion $\sigma$ can be related to the halo parameters $\rho_0$ and $r_0$ via~\cite{Nishikawa:2017chy}
\begin{equation}
\sigma^{2}=\frac{4 \pi G_N \rho_{0} r_{0}^{2} g\left(c_{m}\right)}{c_{m}},
\end{equation}
where $c_{m}=r_{\mathrm{max}}/r_0 \approx 2.16$ corresponds to the radius $r_{\mathrm{max}}$ where the circular velocity is maximized, and $g(x)=\ln(1+x)-\frac{x}{1+x}$.

Third, we introduce the halo concentration parameter, defined as $c(M_{\mathrm{vir}}) \equiv r_{\mathrm{vir}}/r_0$, where $M_{\mathrm{vir}}$ is the mass enclosed within the virial radius $r_{\mathrm{vir}}$ given by
\begin{equation}
	M_{\mathrm{vir}} = \rho_{\mathrm{crit}}\Delta\left(\frac{4 \pi\left(c\left(M_{\mathrm{vir}}\right) r_{0}\right)^{3}}{3}\right)=4 \pi \rho_{0} r_{0}^{3} g\left(c\left(M_{\mathrm{vir}}\right)\right),\label{Mvir}
\end{equation}
where $\Delta \approx 200$ is the overdensity parameter and $\rho_{\mathrm{crit}}$ is the critical density of the Universe.

Fourth, the concentration parameter $c$ is known to correlate with the virial mass $M_{\mathrm{vir}}$ through empirically calibrated relations, often expressed as a polynomial in $\log(M_{\mathrm{vir}})$ as~\cite{Nishikawa:2017chy}
\begin{align}
c\left(M_{\mathrm{vir}}\right) & = \sum_{i=0}^{5} c_i \left[ \log_{10} \left(\frac{h M_{\mathrm{vir}}}{\mathrm{M}_{\odot}}\right) \right]^i,
\end{align}
where $h$ is the dimensionless Hubble parameter, $h_{0}=0.67$ and the coefficients $c_i$ are fitted from simulations, $c_{0}=37.5153, c_{1}=-1.5093, c_{2}= 1.636 \times 10^{-2}, c_{3}=3.66 \times 10^{-4}, c_{4}=-2.89237 \times 10^{-5}$ and $c_{5}=5.32 \times 10^{-7}$.

Using the above equations, we obtain a relation to connect $M$ and $M_{\mathrm{vir}}$ via
\begin{equation}
	\sigma^{2}\left(M\right)=\frac{4 \pi}{3} G_N \Delta \rho_{c} \frac{g\left(c_{\mathrm{m}}\right) c}{g(c) c_{\mathrm{m}}}\left(\frac{3 M_{\mathrm{vir}}}{4 \pi \Delta \rho_{c}}\right)^{2 / 3}.\label{MMvir}
\end{equation}
We can then solve for $M_{\mathrm{vir}}$ numerically for each SMBH mass $M$ using~\eqref{MMvir}, and subsequently obtain $\rho_{0}$ and $r_0$ using~\eqref{Mvir}. With the halo parameters $\rho_0$ and $r_0$ thus determined for a given SMBH mass $M$, the spike parameters $\rho_R$ and $R_{\mathrm{sp}}$ can be calculated using the formulae provided in the Section~\ref{sec:DESR}, allowing for the evaluation of the spike density profile $\rho_{\mathrm{sp}}(r)$.

\section{Numerical details for solving the radial superradiance equation}
\label{app:numerical}

This Appendix details the numerical methods used to solve the Klein-Gordon equation with a position-dependent effective mass, $\mu_{\mathrm{eff}}^2(r)$, as required for Model II (Section~\ref{sec:DESR}). The standard procedure involves separating the scalar field $\Phi(t, r, \theta, \phi)$ using the ansatz $e^{-i \omega t} e^{i m \phi} R(r) S(\theta)$, leading to distinct angular and radial ordinary differential equations.

The angular part results in the standard spheroidal harmonic equation. Transforming the variable $x = \cos \theta$, the equation becomes
\begin{align}
&\left(1-x^{2}\right) \frac{\mathrm{d}^2 S(x)}{\mathrm{d} x^{2}}-2 x \frac{\mathrm{d} S(x)}{\mathrm{d} x} \nn \\
&+\left(\Lambda_{\ell m}+a^{2} \omega^{2} x^{2}-\frac{m^{2}}{1-x^{2}}\right) S(x)=0\,.
\end{align}
The required boundary conditions are the regularity conditions at the poles ($x=\pm 1$) by
\begin{equation}
\lim _{x \rightarrow-1} S(x) \sim(1+x)^{|m| / 2}, \quad \lim _{x \rightarrow+1} S(x) \sim(1-x)^{|m| / 2}\,.
\end{equation}
This equation can be solved numerically using Leaver's continued fraction method~\cite{Leaver:1985ax,Dolan:2007mj,Liu:2022ygf}. By expanding the solution as
\begin{equation}
S(x)=\exp (a \omega x)(1-x)^{|m| / 2}(1+x)^{|m| / 2} \sum_{n=0}^{\infty} b_{n}(1+x)^{n},
\end{equation}
the coefficients $b_n$ are found to satisfy a three-term recurrence relation,
\begin{equation}
\left\{\begin{array}{l}
\alpha_{0} b_{1}+\beta_{0} b_{0}=0, \\
\alpha_{n} b_{n+1}+\beta_{n} b_{n}+\gamma_{n} b_{n-1}=0\,, \quad (n\geq 1)
\end{array}\right.
\end{equation}
where $b_{0}=1$, and the coefficients $\alpha_n, \beta_n, \gamma_n$ are functions of $n, \ell, m, a, \omega$ and the separation constant $\Lambda_{\ell m}$ via
\begin{equation}
\left\{\begin{aligned}
\alpha_{n} &= -2(n+1)(n+|m|+1)\,, \\
\beta_{n} &= n(n+1) + a^{2}\omega^{2} - 2a\omega m + |m|(|m|+1) - \Lambda_{\ell m} \\
&\quad + 2a\omega(n+|m|) - (2n+1)(a\omega+|m|)\,, \\
\gamma_{n} &= 2 a \omega(n+|m|)\,.
\end{aligned}\right.
\end{equation}
For the series solution to converge, the coefficients must satisfy the continued fraction equation,
\begin{equation}
0=\beta_{0}-\displaystyle\frac{\alpha_{0} \gamma_{1}}{\beta_{1}-\displaystyle\frac{\alpha_{1} \gamma_{2}}{\beta_{2}-\displaystyle\frac{\alpha_{2} \gamma_{3}}{\beta_{3}-\ldots}}}\,.
\end{equation}
Given $a$, $\omega$, $\ell$, and $m$, this equation is solved numerically to find the eigenvalue $\Lambda_{\ell m}$.

However, the radial equation Eq.~\eqref{radialeq} is modified by the position-dependent effective mass $\mu_{\mathrm{eff}}^2(r)$ and cannot be solved using the same continued fraction technique. We therefore employ a direct integration method, specifically the ``shooting method''~\cite{Molina:2010fb} to find the complex eigenfrequency $\omega$.

The shooting method requires imposing appropriate boundary conditions near the BH horizon ($r \to r_+$) and at spatial infinity ($r \to \infty$). These are implemented by finding asymptotic series solutions in these limits. Near the horizon and at spatial infinity, we expand the radial function $R(r)$ as
\begin{align}
R^{\mathrm{horizon}}(r) &= (r-r_{+})^{\sigma_{+}}\sum_{i=0}^{\infty} h_{\mathrm{H}}^{(i)}(r-r_{+})^i\,, \\
R^{\infty}(r) &= e^{ik_\infty r}r^{\gamma-1}\sum_{i=0}^{\infty} h_{\infty}^{(i)}r^{-i}\,,
\end{align}
where $\sigma_{+} = i(m\Omega_H - \omega)(r_+^2+a^2)/(r_+-r_-)$, $k_\infty = \sqrt{\omega^2-\mu^2_{\mathrm{eff}}(r\to\infty)}$ (which simplifies as $\mu_0 \approx 0$), and $\gamma$ depends on $\omega, M, \mu$. The condition $\omega_I > 0$ with $\omega_R < m\Omega_H$ for superradiance corresponds physically to requiring purely ingoing waves at the horizon and purely outgoing waves at infinity.

Substituting these series expansions into the radial equation allows us to determine the expansion coefficients $h^{(i)}$ recursively. We then numerically integrate the radial equation inwards from a large radius (``numerical infinity'') and outwards from near the horizon, using the series solutions to provide initial conditions close to the boundaries. The complex eigenfrequency $\omega$ is found by searching for values that allow the inwards and outwards solutions (and their derivatives) to match smoothly at an intermediate radius $r_{\mathrm{M}}$. We verified that our numerical results are insensitive to the choice of $r_{\mathrm{M}}$.

Care must be taken when implementing the numerical integration from spatial infinity. The general solution at large $r$ contains both exponentially decaying (physical) and exponentially growing (unphysical) modes. To find the correct quasi-bound state, we must select the purely decaying solution. Numerical errors can inadvertently introduce contributions from the growing mode, which can dominate the solution if the integration starts from too large a radius (``numerical infinity''). This issue can be mitigated by choosing a sufficiently small value for numerical infinity (e.g., $r_\infty \sim 50M$) and using a sufficiently high order in the asymptotic series expansion to set the initial conditions accurately~\cite{Molina:2010fb}.

%
\bibliography{ref}
\bibliographystyle{utphys}

\end{document}